\documentclass[seceq]{ptptex}

\usepackage{epsfig}

\notypesetlogo                       
\def\lsim{\mathrel{\raise.3ex\hbox{$<$\kern-.75em\lower1ex\hbox{$\sim$}}}}
\def\gsim{\mathrel{\raise.3ex\hbox{$>$\kern-.75em\lower1ex\hbox{$\sim$}}}}

\title{
Big-Bang Nucleosynthesis Reactions  
Catalyzed  by  \\
a Long-Lived Negatively Charged Leptonic Particle
}

\author{
Masayasu KAMIMURA,$^{1,3}$
Yasushi KINO$^{2}$ and Emiko HIYAMA$^{3}$
}

\inst{
$^{1}$Department of Physics, Kyushu University, 
Fukuoka 812-8581, Japan\\
$^{2}$Department of Chemistry, Tohoku University, 
Sendai 980-8578, Japan\\
$^{3}$RIKEN Nishina Center, RIKEN, Wako 351-0198, Japan
}

\abst{
An accurate quantum three-body calculation is performed for 
the new type of big-bang nucleosynthesis (BBN) reactions that are
catalyzed by a hypothetical long-lived 
negatively charged, massive leptonic particle (called $X^-$)
such as the supersymmetric (SUSY) particle {\it stau}, 
the scalar partner of the tau lepton.
It is known that if  the $X^-$ particle has a lifetime  of
$\tau_X \gsim 10^3$ s, it can capture 
a light element previously synthesized in standard BBN and
form  a Coulombic bound state,  for example,
$(^7{\rm Be} X^-)$ at temperature $T_9$  $\lsim 0.4$ 
(in units of $10^9$ K),
$(\alpha X^-)$ at $T_9 \lsim 0.1$ and 
$(p X^-)$ at  $T_9 \lsim 0.01$.
The bound state, an exotic atom, is expected to induce
the following reactions in which $X^-$ acts as a catalyst: 
i) $\alpha$-transfer reactions such as 
$(\alpha X^-)+d  \to~^6{\rm Li} +X$,
ii) radiative capture reactions such as 
$(^7{\rm Be}X^-)+ p \to (^8{\rm B}X^-) + \gamma$,
iii) three-body breakup reactions such as
$(^7{\rm Li}X^-)+ p \to~\alpha + \alpha + X^-$,
iv) charge-exchange reactions such as
$(p X^-)+ \alpha \to~(\alpha X^-) +p$ and 
v) neutron induced reactions such as
$(^8{\rm Be}X^-)+ n \to~\!^9{\rm Be} + X^-$.
In recent papers it has been claimed that 
some of these $X^-$-catalyzed reactions 
have significantly large cross sections so that the
inclusion of the reactions into the BBN network calculation
can markedly change the abundances of some elements, 
giving not only a solution to  the  
$^6$Li-$^7$Li problem (the calculated underproduction of 
$^6$Li by a factor of $\sim 1000$  and  
overproduction of $^7$Li$+$$^7$Be by a factor of $\sim 3$)
but also a constraint on the lifetime and  primordial abundance
of the elementary particle $X^-$.
However, most of these calculations of
the reaction cross sections in the literature 
were performed assuming 
too naive models or approximations that are unsuitable
for these complicated low-energy nuclear reactions.
We use a high-accuracy few-body calculation method
developed by the authors 
and provide precise cross sections and rates of
these catalyzed BBN reactions 
for use in the BBN network calculation.
}

\begin{document}

\maketitle

\section{Introduction}

The accurate theoretical prediction of nuclear reaction rates
is of  essential importance
in the study of big-bang nucleosynthesis (BBN)
when particular reaction rates  are very difficult or even 
impossible to estimate experimentally.
The most typical example of such reactions is the
production and destruction 
of light elements catalyzed by
a hypothetical long-lived negatively charged  
massive ($\gsim 100$ GeV)
leptonic particle, here denoted as $X^-$;
a candidate for this particle 
is the supersymmetric (SUSY) counterpart 
of the tau lepton  $(\tau)$, 
i.e., the stau $(\tilde{\tau}).$\footnote{The stau is   
one of the particles expected to 
be discovered at the CERN Large Hadron Collider 
\cite{LHC2007}.}   
Recently,  BBN involving these $X^-$-{\it catalyzed} 
nuclear reactions has been
extensively studied from the viewpoint of 
particle-physics catalysis in BBN
\cite{Pospelov2007,Kohri2006,Kaplinghat,Cyburt2006,Hamaguchi,
Bird2007,Kawasaki,Jittoh,Jedamzik2007a,Jedamzik2007b,
Kusakabe2007a,Kusakabe2007,Pospelov2007-9Be,Pospelov2008-9Be},
which may be briefly reviewed as follows (and precisely reviewed 
in Refs.~\citen{Kusakabe2007} and \citen{Pospelov2008-9Be} and
references therein): 

\parindent 10pt
i) If  the  $X^-$ particle has a lifetime on the order of $10^3$ s
or longer, it can capture a light nucleus, $A$, 
previously synthesized during BBN, and produce 
a bound state denoted  as $(AX^-)$, which is 
a type of exotic atom with the nucleus $A$ 
in a Coulombic orbit around 
the massive $X^-$ at 
the center.\footnote{For instance,     
the binding energies of $(^7{\rm Be} X^-)$ and  $(\alpha X^-)$ 
are  $\sim$1.3  and  $\sim$0.34 MeV, 
and  the rms radii are   
$\sim$3.6 and $\sim$7 fm, respectively.}

\parindent 10pt
ii) When the cosmic temperature cools to 
$T_9 \sim 0.4$ (in units of $10^9$~K) after the standard
BBN $(T_9 \gsim 0.5)$, $X^-$  captures 
$^7$Be to form $(^7{\rm Be} X^-)$, which is subsequently 
destroyed by the $(^7{\rm Be} X^-) + p \to 
(^8{\rm B} X^-) + \gamma$ reaction\cite{Bird2007}
followed by the $\beta$-decay of $(^8{\rm B} X^-)$ to
$\alpha + \alpha +X^-$.

\parindent 10pt
iii) At $T_9 \sim 0.3$, $X^-$ captures
$^7$Li to generate $(^7{\rm Li} X^-)$,
which is subsequently 
destroyed by the $(^7{\rm Li} X^-) + p \to 
\alpha + \alpha +X^-$ reaction.

\parindent 10pt
iv) At $T_9 \sim 0.1$, $X^-$ particles  capture  
$\alpha$ particles to form  $(\alpha X^-)$ in abundance
which triggers the reaction
$(\alpha X^-) + d \to~^6{\rm Li} + X^- $ to produce
an enormous amount of $^6$Li, as 
originally proposed by Pospelov\cite{Pospelov2007} and
confirmed by Hamaguchi {\it et al.}\cite{Hamaguchi}
including two of the present authors (M.~K. and Y.~K.).

\parindent  10pt
v) At $T_9 \sim 0.01$, although $X^-$ particles start capturing 
protons to form $(p X^-)$, its abundance remains low due to  
the charge-exchange 
reaction $(p X^-) + \alpha \to (\alpha X^-) + p$,  
which occurs before $(p X^-)$ begins to interact 
with other nuclei to form heavier 
elements.\cite{Pospelov2008-9Be}

\parindent 20pt
The contribution of these reactions 
is expected to modify standard BBN (SBBN in short) and
to solve the calculated underproduction 
(by a factor of $\sim$1000) of the  
primordial abundance of $^6$Li and the
overproduction (by a factor of $\sim$3) of $^7$Li \cite{6Li7Li},
and, at the same time,
to provide information for constraining
the lifetime and primordial 
abundance of the elementary particle $X^-$.
However, full quantum mechanical studies of the reactions 
have not been performed
on the basis of nuclear reaction theory
except for the work of Ref.~\citen{Hamaguchi} 
on the production of $^6$Li.

The purpose of the present paper is to perform 
a precise quantum three-body calculation of
the cross sections of many types of catalyzed BBN (CBBN in short) 
reactions including those 
mentioned above in i) -- v) and to provide 
their reaction rates for 
use in the BBN network calculation.
The adopted theoretical method and numerical 
technique have been developed by the present authors, 
reviewed in Ref.~\citen{Hiyama2003},
and are well established in the field of
few-body atomic and nuclear systems.  

The CBBN reactions studied in this paper are classified 
into four types:

\noindent
1) {\it Transfer reactions}:
\begin{eqnarray}
(AX^-) + a \to B + X^- ,
\end{eqnarray}
\hskip 0.5cm where $A$ is picked up by an incoming nucleus $a$ 
to produce a nucleus $B (=A+a)$,

\noindent
2) {\it Radiative capture reactions}:
\begin{eqnarray}
(AX^-) + a \to (B X^-) + \gamma \, ,
\end{eqnarray}

\noindent
3) {\it Three-body breakup reactions}:
\begin{eqnarray}
(AX^-) + a \to b_1 + b_2  + X^- ,
\end{eqnarray}

\noindent
4) {\it Charge-exchange reactions}:
\begin{eqnarray}
(AX^-) + a \to (a X^-) + A .
\end{eqnarray}
We solve the Schr\"{o}dinger equations
of CBBN reactions (1$\cdot$1)$-$(1$\cdot$4), 
explicitly taking the three-body 
degree of freedom into account,
and derive their cross sections and reaction rates
at incoming energies below 200  keV.

Throughout this paper, it is not necessary to identify $X^-$ 
with any particular particle such as the SUSY particle stau,
although this is assumed, for instance,
in Refs.~\citen{Cyburt2006}, \citen{Hamaguchi}
and \citen{Pospelov2008-9Be}. 
The property of $X^-$ we assume here
is that it has a charge of $-e$, a mass $m_X \gsim 100$ GeV  
and a lifetime $\tau_X \gsim 10^3$ s and that
it interacts with light elements via the Coulomb force only.

The structure of the present paper is  as follows.
We investigate  $X^-$-catalyzed $\alpha$-transfer reactions
that produce $^6$Li, $^7$Li and $^7$Be in \S 2,   
resonant and nonresonant
 radiative capture reactions that destroy $^7$Be in \S 3,
 three-body breakup reactions that destroy
$^6$Li and $^7$Li in \S 4,
the charge-exchange reactions in \S 5 
and the possibility of the
$X^-$-catalyzed primordial production of $^9$Be in 
\S 6.
A summary is given in \S 7.



\section{$X^-$-catalyzed $\alpha$-transfer reactions}


\subsection{Necessity of three-body calculation}

\subsubsection{Production of $^6$Li}

One of the puzzles in SBBN is that 
the BBN prediction of the $^6$Li abundance is a factor of
$\sim$1000 smaller than the observed data\cite{6Li7Li}. 
The too small prediction is mostly because the radiative capture 
reaction that produces $^6$Li in SBBN, 
\begin{eqnarray}
\alpha+d \to~^6{\rm Li}+ \gamma \,( \geq 1.47 \, {\rm MeV}) ,
\end{eqnarray} 
is  heavily E1-hindered with
a very small astrophysical $S$-factor, 
$S_\gamma \sim 2 \times 10^{-6}$ keV~b\cite{Angulo1999}.
To obtain a much greater production of $^6$Li,
Pospelov considered \cite{Pospelov2007}, for the first time,
a new type of CBBN reaction,
\begin{eqnarray}
(\alpha X^-) + d \to~^6{\rm Li}+X^-  + 1.13 \,{\rm MeV},
\label{eq:CBBN_6Li}
\end{eqnarray}
in which the bound state\footnote{Hereafter,
when $(AX^-)$ is in the 
$1s$ ground state, the suffix $1s$ of $(AX^-)_{1s}$ 
is omitted.} 
$(\alpha X^-)$ promotes the synthesis of
$\alpha$ and $d$ to form $^6$Li with no photon emission.
He claimed that the $S$-factor of the $X^-$-catalyzed reaction 
(2$\cdot$2), say $S_X$, is enhanced by 
8 orders of magnitude
$(S_X \sim 3 \times 10^2$ keV~b)
compared with $S_\gamma$
and that the abundance of $^6$Li is greatly increased 
so that the $^6$Li problem  can be solved
under a suitable 
restriction on the abundance and lifetime of $X^-$.

His estimation of 
$S_X$, however, was based on a too naive comparison between 
(2$\cdot$1) and (2$\cdot$2).
Namely,  in an analogy to the real photon in (2$\cdot$1)
with the wavelength $\lambda_\gamma \sim 130 $ fm,
Pospelov introduced a virtual photon in 
(2$\cdot$2) with a characteristic 
wavelength $\lambda_X$ on the order of
the Bohr radius ($=3.63$ fm) of the atom $(\alpha X^-)$ and assumed
the scaling relation\footnote{A more detailed expression 
is given by Eq.~(4) in Ref.~\citen{Pospelov2007}
but the scaling relation is  essentially described by (2$\cdot$3).}
\begin{eqnarray}
  \frac{ S_X}{S_\gamma} \sim 
\left( \frac{\lambda_\gamma}{\lambda_X}\right)^{2\ell+1}, 
\end{eqnarray}
where the photon multipolarity $\ell$ is assumed 
to be $\ell=2$ (namely, E2), not only for 
the real photon\footnote{This            
assumption of $\ell=2$ for the real photon  in (2$\cdot$1)
may not be justified  {\it at astrophysical energies}
(no experimental data there).
According to the so far most precise six-body 
calculation \cite{NWS01} of the E1-hindered 
radiative capture reaction (2$\cdot$1), 
the $S$-factor of the E2 transition decreases much more rapidly than
that of the E1 transition as the energy decreases, and eventually
the E1/E2 ratio becomes $\sim 10$ 
at $E \sim 0$ (see Figs.~7 and 8 in Ref.~\citen{NWS01}).}
but also for the virtual one. In this case,
the enhancement ratio $S_X/S_\gamma$ amounts to $\sim 10^8$.

However, the scaling 
relation (2$\cdot$3), which compares the wavelengths of 
the real and virtual photons, is not suitable
for discussing reaction (2$\cdot$2) because
the relation does not take into account the  
nuclear interaction, which is the most important factor
in the process of transferring the 
$\alpha$ particle from $(\alpha X^-)$  
to the deuteron to form  $^6$Li.
Therefore, two of the authors (M.~K. and Y.~K.) and
their colleagues \cite{Hamaguchi}
performed a fully quantum three-body calculation
of the $\alpha+d+X^-$ system 
and calculated the $S$-factor of reaction (2$\cdot$2), $S_X(E)$,
as a function of
the incident energy $E$ at $E=10 - 120 $ keV. 
The calculated $S$-factor at the Gamow peak energy
is 38 keV~b ($S_X/S_\gamma \sim 10^7$, 
nearly one order of magnitude smaller than 
that in Ref.~\citen{Pospelov2007}).
Note that  $S_X$ of the CBBN reaction (2$\cdot$2)
is of a similar order of magnitude to the    
$S$-factors of  nonresonant photonless SBBN reactions 
caused by the nuclear interaction;
for example, compare  $S_X$ with the observed values of
$\sim 60$ keV~b for $d+d \to t + p$
and $\sim 5 \times 10^{-6}$ keV~b for the E1-forbidden process
$d+d \to \alpha + \gamma$.
The large enhancement of the CBBN/SBBN ratio $(\sim 10^7)$ 
is simply because the SBBN reaction (2$\cdot$1) 
is heavily E1-hindered.

\subsubsection{Production of $^7$Li and $^7$Be}
In the BBN network calculation with 
the CMB-based  baryon-to-photon ratio 
($\eta_{\rm B}=6.0 \times 10^{-10}$),\cite{CMB}
the abundance of $^7{\rm Be}$ is 
approximately one order of magnitude
larger than that of  $^7{\rm Li}$. 
Since $^7{\rm Be}$ eventually 
decays to  $^7{\rm Li}$ by electron capture,
the sum of the abundances of both $^7$Li and $^7$Be
is regarded as the abundance of $^7{\rm Li}$ in the 
$^7{\rm Li}$-overproduction  problem.
To examine the effect of CBBN on the
abundance of  $^7{\rm Li}$+$^7{\rm Be}$,  
Cyburt {\it et al.} \cite{Cyburt2006} studied 
the $\alpha$-transfer reactions
\begin{eqnarray}
&&(\alpha X^-)+t \to~^7{\rm Li}+X^-  + 2.13 \, {\rm MeV} ,
\\
&&(\alpha X^-)+\!\!~^3{\rm He} \to~^7{\rm Be}+X^-  
+ 1.25 \, {\rm MeV} .
\end{eqnarray}
which produce $^7$Li and $^7$Be at $T_9 \sim 0.1$.
Using the scaling relation proposed in Ref.~\citen{Pospelov2007}, 
they claimed that the $S$-factors of the above CBBN reactions
are  5 orders of magnitude  greater
than those of the E1 radiative capture reactions
\begin{eqnarray}
&&\alpha + t \to~^7{\rm Li} + \gamma \,
( \geq 2.47 \, {\rm MeV}) \: ,
\\
&&\alpha +\!\!~^3{\rm He} \to~^7{\rm Be}+ \gamma 
\, ( \geq 1.59 \, {\rm MeV}) ,
\end{eqnarray}
which are the most effective source for  producing 
$^7$Li and $^7$Be in SBBN.

However, it is clear that the enhancement ratio
CBBN/SBBN in $S$-factors will be only on the order of $10^1 - 10^2$ 
contrary to the case of $^6$Li production ($\sim 10^7$).
The reason for this is as follows: the SBBN
$S$-factors of (2$\cdot$6) and (2$\cdot$7) at astrophysical energies are
respectively $\sim 0.1$ and $\sim 0.5$ keV~b, \cite{Angulo1999} 
which are 5 orders of magnitude larger than 
that of the E1-hindered reaction (2$\cdot$1).
Moreover,  the CBBN $S$-factors
of (2$\cdot$4) and (2$\cdot$5) should be significantly reduced
compared with that of (2$\cdot$2), because a
rearrangement of the angular momenta 
among the  three particles is needed when going from 
the entrance channel to the exit channel; note that 
a significant difference between (2$\cdot$2) and
(2$\cdot$4)$-$(2$\cdot$5) is that the relative motion between 
$d$ and $\alpha$ in $^6$Li is that of $s$-wave whereas that 
between $t (^3{\rm He})$ and $\alpha$ in $^7$Li ($^7$Be) is
that of  $p$-wave. 
This consideration of the CBBN/SBBN ratio
will be confirmed by the quantum 
three-body calculation of reactions 
(2$\cdot$4) and (2$\cdot$5) in this section.

\subsection{Three-body Schr\"{o}dinger equation and reaction rate}

Following Refs.~\citen{Hamaguchi} and \citen{Hiyama2003}, 
we briefly
explain the three-body calculation  method used
to investigate the above $X^-$-catalyzed transfer reactions.
We emphasize that 
the method described here is an exact method
for calculating the cross section of the low-energy transfer
reactions 
within the three-body model of the total system concerned.
The present authors have previously  
performed the same types of calculations
in the study of muon transfer reactions \cite{
Kamimura1988a,Kamimura1988b,Kino1993,Hiyama2003}
in muon-catalyzed fusion cycles \cite{Nagamine1998}.

As shown in Fig.~1, we consider all three sets of the
 Jacobi coordinates $({\boldsymbol r}_c, {\boldsymbol R}_c)$,
$c=1,2$ and $3$,
to completely treat the three-body degrees of freedom 
of the system $A+a+X^-$.
Here, we take $A=\alpha$, and $a=d, t$ and $^3$He
for reactions (2$\cdot$2), (2$\cdot$4) and (2$\cdot$5), respectively.

\begin{figure}[bth]
\begin{center}
\epsfig{file=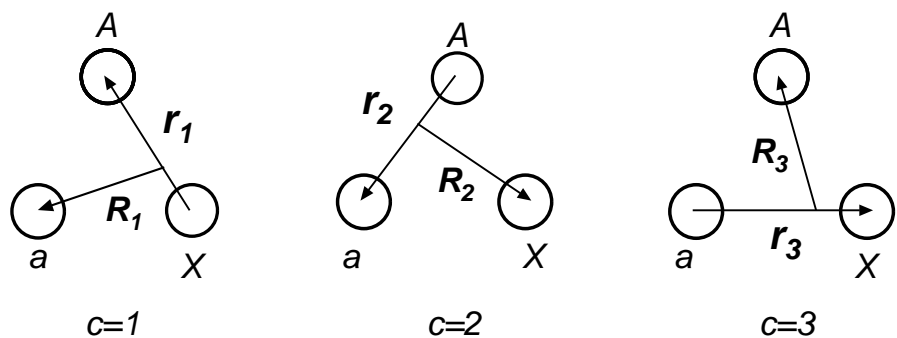,width=12cm,height=4.5cm}
\end{center}
\caption[]{ 
Three sets of Jacobi coordinates in
the $ A+a+X^-$ system. The entrance  
 channel $(AX^-) + a$ is described  using the coordinate 
system  $c=1$, and the transfer channels $(Aa) + X^-$  and 
$(aX^-) + A$ are described using $c=2$ and $c=3$, respectvely.
}
\label{fig:jacobi}
\end{figure}


The Schr\"{o}dinger equation for  
 the total wave function 
with total energy $E_{\rm tot}$,  
angular momentum $J$ and  $z$-component $M$
is given by 
\begin{equation}
 ( H - E_{\rm tot} ) \Psi_{JM} = 0   
\label{eq:schroedinger}
\end{equation}
with the Hamiltonian
\begin{eqnarray}
H=-{\hbar^2 \over 2m_{c}} \nabla^{2}_{{\boldsymbol r}_{c}}
           -{\hbar^2 \over 2M_{c}} \nabla^{2}_{{\boldsymbol R}_{c}}
          + V_{A{\mbox -}X}(r_1) 
          + V_{A{\mbox -}a}(r_2) 
          + V_{a{\mbox -}X}(r_3) . \qquad 
\label{eq:hamiltonian}
\end{eqnarray}
As far as we use the reduced masses
 ($m_c$ and $M_c$)  associated with the coordinates 
 (${\boldsymbol r}_c$ and ${\boldsymbol R}_c$),  
every choice of $c$ in the 
  kinetic term is equivalent. The notation of the potentials is 
self-evident; an explicit form will be given below. 
The Schr\"{o}dinger equation is solved 
under the scattering boundary condition
imposed appropriately on the relevant reaction.

For all the CBBN reactions studied in this paper,
only the resonant radiative capture reaction (1$\cdot$2) requires
serious consideration of the intrinsic spins of the particles 
as will be discussed in \S 3.
In this section, for the nonresonant $\alpha$-transfer reactions,  
the spin of the incoming particle 
$a~(=d, t, ^3\!{\rm He}$) is neglected.
Therefore, in the exit channel of (2$\cdot$4) and (2$\cdot$5), 
 the $3/2^-$($p$-wave) ground state 
and the $1/2^-$($p$-wave)
first excited state of $^7$Li and
$^7$Be  are degenerated energetically to their weighted mean. 
The validity of this approximation will be
discussed in \S 2.6.

Firstly, we construct  the wave function of  the  $1s$ ground state 
of  $(AX^-)$   in the entrance channel
on the coordinate system $c=1$
and that of the   ground state of
the nucleus $B$ on $c=2$.
We denote  the wave function by $\phi_{l_c m_c}^{(c)}({\boldsymbol r}_c)$
and the eigenenergy by  $\varepsilon_{l_c}^{(c)}$;
explicitly, $l_1=0$  for  $(\alpha X^-)$,
$l_2=0$ for $^6$Li and $l_2=1$ for $^7$Li ($^7$Be). 
These quantities are obtained by solving
\begin{equation}
  \big[  -{\hbar^2 \over 2m_{c}} \nabla^{2}_{{\boldsymbol r}_{c}}
     +   V_c(r_c) - 
\varepsilon_{l_c}^{(c)} \,\big] \,
   \phi_{l_c m_c}^{(c)}({\boldsymbol r}_c)=0 ,
 \quad (c=1,2) 
\label{eq:subsystem_c}
\end{equation}
where the potential $V_c$ denotes  $V_{A{\mbox -}X}$ and 
$V_{A{\mbox -}a}$ for $c=1$ and 2, respectively.

Using 
$\chi^{(c)}_{L_c M_c}({\boldsymbol R}_c) \equiv \chi^{(c)}_{L_c}(R_c)
 Y_{L_c M_c}({\widehat {\boldsymbol R}}_c)$, we denote 
the scattering wave function
along the coordinate ${\boldsymbol R}_c$ 
with angular momentum $L_c$ and 
$z$-component $M_c$ $\;(c=1,2)$. 
The center-of-mass (cm) scattering energy  
associated with the coordinate ${\boldsymbol R}_c$, say $E_c$, 
is introduced as
\begin{equation}
E_c= E_{\rm tot} - \varepsilon_{l_c}^{(c)}=\hbar^2 k^2_c /2M_c, 
\quad(c=1,2)
\end{equation}
together with the corresponding wave number $k_c$.

The total three-body wave function $\Psi_{JM}$,
which describes the transfer reaction
(2$\cdot$2) in the most sophisticated manner, is written as
\cite{Kamimura1977,Hiyama2003} 
\begin{equation}
\Psi_{JM}=\phi^{(1)}_{00}({\boldsymbol r}_1)\,
\chi^{(1)}_{JM}({\boldsymbol R}_1)   +
\left[\phi^{(2)}_{l_2}({\boldsymbol r}_2)\otimes 
\chi^{(2)}_{L_2}({\boldsymbol R}_2)\right]_{JM}
         +\Psi^{({\rm closed})}_{JM}\;.
\label{eq:psi12}
\end{equation}
Here, the first and second terms represent 
the two open channels, $(AX^-)+a$ on $c=1$ and
$ B+X^-$ on $c=2$, respectively; there are no other open channels
in the energy range $(E_1< 150 $ keV) that we  are interested in. 
In the first term, 
the radial part 
$\chi^{(1)}_{J}(R_1) $ 
has incoming and outgoing amplitudes,
and $\chi^{(2)}_{L_2}(R_2)$ in the second term
has the outgoing amplitude only.
These should satisfy the boundary condition
\begin{eqnarray}
\lim_{R_{c}\to\infty} R_c \chi^{(c)}_{L_c}(R_c) = 
u^{(-)}_{L_c}(k_{c},R_{c})\delta_{c \,1}
 -\sqrt{\frac{v_1}{v_{c}}}
S^{L_c}_{1 \to c}u^{(+)}_{L_c}(k_{c},R_{c}) ,
\qquad (c=1,2) \qquad
\label{eq:boundary}
\end{eqnarray}
where $L_1=J$, and 
$u^{(\pm)}_{L}(k_cR_c)(=G_L(k_cR_c) \pm i F_L (k_cR_c))$  
are the asymptotic outgoing and incoming
Coulomb wave functions. 
$S^{L_c}_{1 \to c }$ is  the $S$-matrix for the transition
from  the incoming channel $(c=1)$ to the outgoing channel $c$ and 
$v_c$ is the velocity. 

The third term in (\ref{eq:psi12}), $\Psi^{({\rm closed})}_{JM}$,
represents all the closed (virtually excited) channels 
in the energy range in this work; 
in other words, this term is responsible for
all the asymptotically vanishing amplitudes due to the 
three-body degrees of freedom
that are not included in the first 
two scattering terms.\footnote{              
This method for low-energy three-body reactions
has already been used \cite{Kamimura1988a,
Kamimura1988b,Kino1993,Hiyama2003}
in the study of  
the muon transfer reaction
$(d\mu)_{1s} + t \to (t\mu)_{1s} + d + 48\, {\rm eV}$.   
The  $\Psi^{({\rm closed})}_{JM}$ term 
was found to play a very important role; 
if the term is omitted (namely, if the two-channel coupled
calculation is performed with $c=1,2$),
the calculated low-energy ($0.001 -100$ eV)
cross section of the reaction
becomes $\sim$ 30 times larger 
than that obtained by the full three-body
calculation.}
Since $\Psi^{({\rm closed})}_{JM}$
vanishes asymptotically, it is  reasonable
and useful  
to expand it in terms of a 
complete set of  $L^2$-integrable three-body basis functions, 
$\{ \Phi_{JM, \nu}; \nu=1-\nu_{\rm max} \}$,
that are spanned in a finite spatial region (see \S 2.4):
\begin{equation}
\Psi^{({\rm closed})}_{JM}  =  \sum_{\nu=1}^{\nu_{\rm max}}
  b_{J\,\nu} \; \Phi_{JM,\:\nu}     .
\label{eq:Phi-def}
\end{equation}
Equations for $\chi^{(1)}_{J}(R_1)$, 
$\chi^{(2)}_{L_2}(R_2)$ 
and the coefficients $b_{J \nu}$ are given by the $\nu_{\rm max}+2$
simultaneous equations \cite{Kamimura1977,Hiyama2003},
\begin{eqnarray}
\langle \, \phi^{(1)}_{\,0 0}({\boldsymbol r}_1) 
  Y_{JM}(\widehat{\boldsymbol R}_1)\,
  \:   | \,H - E_{\rm tot} \,|\:
\Psi_{JM} \,\rangle _{{\boldsymbol r}_1,\,
 \widehat{\boldsymbol R}_1 }&=&0 ,\\
\langle \, \left[ \phi^{(2)}_{\,l_2}({\boldsymbol r}_2) \otimes
  Y_{L_2}(\widehat{\boldsymbol R}_2) \right]_{JM} 
     | \,H - E_{\rm tot} \,|\:
\Psi_{JM} \,\rangle _{{\boldsymbol r}_2,\, 
\widehat{\boldsymbol R}_2 }&=&0 ,
\label{eq:eq27}
\end{eqnarray}
and
\begin{equation}
\langle \,\Phi_{JM,\,\nu}\:|\, H - E_{\rm tot} \, 
| \:\Psi_{JM} \,\rangle =0 .
    \hskip 20 pt  (\nu=1 - \nu_{\rm max})  
\label{eq:eq27}
\end{equation}
Here, $\langle \hskip 3 ex \rangle_{{\boldsymbol r}_c,\,
 \widehat{\boldsymbol R}_c}$
 denotes  integration over 
${\boldsymbol r}_{c}$ and $\widehat{\boldsymbol R}_c$.

Since $\Phi_{JM,\:\nu}$ are constructed 
so as to diagonalize the three-body Hamiltonian as (cf. \S 2.4)
\begin{equation}
\langle \Phi_{JM,\,\nu}| \,  H \,  |\Phi_{JM,\,\nu'}\rangle
           =  E_{J\,\nu} \delta_{\nu \nu'} ,
  \qquad  (\nu,\nu'=1 - \nu_{\rm max}) 
\label{eq:diag}
\end{equation}
the coefficients $b_{J \nu}$ can be written, from Eq.~(\ref{eq:eq27}),
as
\begin{eqnarray}
b_{J \nu}=\frac{-1}{E_{J \nu}-E_{\rm tot}} 
 \langle \Phi_{JM,\, \nu}\,|\, H-E_{\rm tot} \,|\,
  \phi^{(1)}_{ 0 0 }({\boldsymbol r}_1)\chi^{(1)}_{JM}({\boldsymbol R}_1)
 + \left[ \phi^{(2)}_{l_2}({\boldsymbol r}_2) \otimes 
\chi^{(2)}_{L_2}({\boldsymbol R}_2) \right]_{JM}
\:\rangle.  \nonumber \\ (\nu=1 - \nu_{\rm max}) \qquad \qquad
\label{eq:bjnu}
\end{eqnarray}
Using Eqs.~(2$\cdot$15), (2$\cdot$16) and (\ref{eq:bjnu}), 
we reach two coupled integro-differential
equations for $\chi^{(1)}_{J}(R_1)$ 
and $\chi^{(2)}_{J}(R_2)$, which are  
not written here 
 (see \S 8 of Ref.~\citen{Hiyama2003} for the equations).

Finally, the integro-differential equations
are solved 
using both the direct numerical method (the finite-difference method)
and the Kohn-type variational method
\cite{Kamimura1977,Hiyama2003}.  We  obtained
the same result;
this demonstrates the high accuracy of 
our three-body calculation.

Using the $S$-matrix elements obtained above,
the cross section of the rearrangement process 
is expressed as 
\begin{equation}
 \sigma(E) 
=\frac{\pi}{k_1^2}\sum_{J=0}^{\infty} (2J+1)
        \bigl|S^{L_2}_{1 \to 2} \bigr|^2  .
\label{eq:cross}
\end{equation}
where we introduce a simplified notation, $E\equiv E_1$,
for the energy of the entrance channel.
The astrophysical $S$-factor is derived from 
\begin{equation}
\sigma(E) = S(E)  \, {\rm exp}(-2\pi \eta(E))/ E .
\label{eq:astro}
\end{equation}
Here, ${\rm exp}(-2\pi \eta(E))$ is the Coulomb barrier 
penetration probability, where
$\eta(E)$ is the Sommerfeld parameter of the entrance
channel $(AX^-)+a$, defined  as
\begin{equation}
  \eta(E)= Z_a Z_{AX} e^2 /\hbar v_1,
\end{equation}
where $Z_a e$ and $Z_{AX} e$ are the charges of 
$a$ and $(AX^-)$, respectively.

The reaction rate $\langle \, \sigma v \,\rangle$ at  temperature $T$
is expressed as (cf. Eq.~(4-44) of 
Ref.~\citen{Clayton1983})
\begin{eqnarray}
 \langle \, \sigma v \,\rangle & =& 
\left( \frac{8}{\pi M_1}\right)^{\frac{1}{2}} 
\frac{1}{(kT)^{\frac{3}{2}}}
\int_0^\infty \,S(E) \,
{\rm exp}\left({-\frac{E}{kT}}- 2\pi \eta(E) \right) dE \:,
\end{eqnarray}
where  $k$ is the Boltzmann constant.
If $S(E)$ is simulated
by a linear function of $E$  around the Gamow peak 
energy $E_{\rm 0}$ as 
\begin{eqnarray}
S(E) &\simeq& S(E_{\rm 0}) + 
\left(\frac{\partial S}{\partial E} 
\right)_{\!\!E_0}\,(E - E_{\rm 0} ), \\
E_0 &=& 122.0 \,(Z_a^2 Z_{AX}^2 \mu)^\frac{1}{3}\, T_9^\frac{2}{3}
\quad {\rm keV},
\end{eqnarray}
the reaction rate $N_A \,\langle \, \sigma v \,\rangle$
is expressed, using Eqs.~(4-56) and (4-75) of 
Ref.~\citen{Clayton1983},  as
\begin{eqnarray}
N_A \,\langle \, \sigma v \,\rangle & =& 
7.82  \times 10^6 \,\left( \frac{Z_a Z_{AX}}{\mu } 
\right)^{\frac{1}{3}}  \left[\, S(E_0)   + 
71.8  \left(\frac{\partial S}{\partial E} 
\right)_{\!\!E_0}\!\!
 T_9 \, \right]  \nonumber \\
&\times& \, T_9^{-\frac{2}{3}} 
 \, {\rm exp} \Bigg[ -4.248\, (Z_a^2 Z_{AX}^2 \mu)^{\frac{1}{3}}
 \,T_9^{-\frac{1}{3}} \Bigg] \qquad 
{\rm cm}^3 \,{\rm s}^{-1} \, {\rm mol}^{-1},
\label{eq:sigmav}
\end{eqnarray}
where $S(E_{\rm 0})$ and 
$({\partial S}/{\partial E} )_{E_0}$ are
given in units 
of keV$\!$~b and b, respectively, $N_A$ is  
Avogadro's number and
$\mu$ is $M_1$ in units of amu.
As will be seen for any nonresonant reaction
in this paper, we can regard 
$({\partial S}/{\partial E} )_{E_0} \approx {\rm constant} 
\equiv  \alpha$: 
\begin{equation}
  S(E)= S(E_0) + \alpha\,(E-E_0) = S(0) + \alpha E ,
\end{equation}
where $S(0)$ and $\alpha$ are given in units of keV~b and b,
respectively, and  $E$ is in keV.
We then have 
\begin{eqnarray}
N_A \,\langle \, \sigma v \,\rangle & = &
7.82  \times 10^6 \,\left( \frac{Z_a Z_{AX}}{\mu } 
\right)^{\frac{1}{3}}   S(0)   
 \, T_9^{-\frac{2}{3}} 
 \, {\rm exp} \Bigg[ -4.248\, (Z_a^2 Z_{AX}^2 \mu)^{\frac{1}{3}}
 \,T_9^{-\frac{1}{3}} \Bigg] \;\;  \nonumber \\
& \times & 
\Bigg[   1 + \frac{\alpha}{S(0)}
\left\{ 122.0 \, (Z_a^2 Z_{AX}^2 \mu)^\frac{1}{3}\, T_9^\frac{2}{3}
+ 71.8 \, T_9 \right\} \Bigg]  
  {\rm cm}^3 \,{\rm s}^{-1} {\rm mol}^{-1} \, . 
\end{eqnarray}

\subsection{Nuclear and Coulomb potentials}

It is essential in the three-body calculation 
to employ appropriate interactions among the three particles.
Here, we determine the potentials 
$V_{A{\mbox -}X}(r_1), V_{A{\mbox -}a}(r_2)$ 
and $V_{a{\mbox -}X}(r_3)$ in (2$\cdot$9), where $A=~\!\!\alpha$ and
$a=d, t$ and $^3{\rm He}$. 

We assume the Gaussian charge distribution 
$Z e(\pi b^2)^{-3/2} {\rm e}^{-(r/b)^2}$
for $\alpha$, $d$,  $t$ and $^3{\rm He}$;   
here, $Ze$ is the charge and $b=\sqrt{2/3}\, r_0$,
where $r_0$ is  the observed rms charge radius, which is
given by $r_0=1.67$,
2.14, 1.70 and 1.95 fm, respectively.  

The Coulomb potential between $A$ and $X^-$ is  given by 
\begin{equation}
V_{A{\mbox -}X}(r)= -Z_A e^2 \, \frac{{\rm erf}(r/b_A)}{r} ,
\end{equation}
where ${\rm erf}(x)=\frac{2}{\sqrt{\pi}} \int_0^x e^{-t^2} {\rm d}t$ 
is the error function;
and the Coulomb potential is similarly defined
for $V_{a{\mbox -}X}(r)$.
The energy of the $(\alpha X^-)_{1s}$ state is 
$ \varepsilon_{\rm gs}^{(1)}=-337.3$ keV ($-347.6$ keV)
and the rms radius is 6.84 fm ($6.69$ fm)
for $m_X = 100$ GeV ($m_X \to \infty$).

The potential $V_{A{\mbox -}a}(r)$ 
is a sum of the nuclear potential,
$V_{A{\mbox -}a}^{\rm N}(r)$, 
and the Coulomb potential, $V_{A{\mbox -}a}^{\rm C}(r)$.
The latter is given by
\begin{equation}
V_{A{\mbox -}a}^{\rm C}(r)= Z_A Z_a \, e^2
 \, \frac{{\rm erf}(r/\sqrt{b_A^2+b_a^2}\:)}{r} . 
\end{equation}
The nuclear potential $V_{A{\mbox -}a}^{\rm N}(r) $ 
is assumed to have a two-range Gaussian shape as
\begin{equation}
V_{A{\mbox -}a}^{\rm N}(r) = 
v_1 \,e^{-(r/a_1)^2} + v_2 \,e^{-(r/a_2)^2}. 
\end{equation}

i) $\alpha$-$d$ {\it potential}

\noindent
We take
$a_1=0.9$ fm, $v_1=500.0$ MeV, $a_2=2.0$ fm and $v_2=-64.06$ MeV
\cite{Hamaguchi}.
The first term, a repulsive core, is introduced to
simulate the Pauli exclusion principle
that nucleons in the incoming 
deuteron should not occupy the nucleon $s$-orbit in
the $\alpha$ particle during the reaction process
(for this role of the
Pauli principle,\footnote{Use                   
of the sophisticated
orthogonality-condition model (OCM) \cite{Saito} 
for three-body systems is
not necessary in the present case, 
since the Pauli principle  only applies
between $A$ and $a$ among the three particles, 
and we are not treating their compact
bound states.
However, the introduction of the inner repulsive potential is useful
in this type of three-body scattering calculation
to  automatically prevent the
unphysical Pauli forbidden amplitude
from entering the total wave function. }
see, for example, Ref.~\citen{Ikeda1980}).   
The parameters are  determined so that
the solution to the Schr\"{o}dinger equation 
(2$\cdot$10) for $c=2$  reproduces 
observed values of both the energy $\varepsilon_{\rm gs}^{(2)}=
-1.474$ MeV and the rms charge radius 
$2.54$ fm \cite{Tanihata1985} of the ground state of $^6$Li.
Furthermore, the charge density of $^6$Li reproduces
the observed  charge form factor 
of the electron scattering from $^6{\rm Li}$ 
(see  Fig.~2 of Ref.~\citen{Hamaguchi}).
Simultaneously, the use of the potential $V_{\alpha{\mbox -}d}(r_2)$ 
explains the low-energy $s$-wave phase shifts 
of the $\alpha+d$ scattering (see Fig.~2  of Ref.~\citen{Hamaguchi}).

\vskip 0.2cm
ii) $\alpha$-$t$ {\it potential}

\noindent
The nuclear potential between $\alpha$ and $t$ 
is assumed to be  parity dependent;
$a_1=1.0$ fm, $v_1=500.0$ MeV, $a_2=2.7$ fm and $v_2=-46.22$ MeV
for odd angular-momentum states and
$a_1=1.5$ fm, $v_1=500.0$ MeV, $a_2=2.4$ fm and $v_2=-7.0$ MeV
for even angular-momentum states.
The repulsive term 
is introduced
so as to prevent the 
Pauli forbidden states
from entering the total wave function.
The use of the potential reproduces 
the observed energy $\varepsilon_{\rm gs}^{(2)}=
-2.308$ MeV (weighted average for the $3/2^-$ ground state
and the $1/2^-$ excited state)
and the rms charge radius $2.43$ fm
\cite{Tanihata1985}
of the ground state of $^7$Li as well as the
observed values of the low-energy $\alpha$-$t$ scattering  
phase shifts for  the $p$-wave  
(average of the $p_{3/2}$ and $p_{1/2}$ states)
and the $s$-wave.

\vskip 0.2cm
iii) $\alpha$-$^3{\rm He}$ {\it potential}

\noindent 
The nuclear $\alpha$-$^3{\rm He}$ potential
is assumed to have the same shape and parameters as the 
$\alpha$-$t$ potential  except that 
$v_2=-44.84$ MeV for odd states, which reproduces the 
observed weighted mean energy ($\varepsilon_{\rm gs}^{(2)}=
-1.444$ MeV) of the ground state
and the $1/2^-$ excited state.

\subsection{Three-body basis functions}

The $L^2$-integrable three-body basis functions
$\{ \Phi_{JM, \nu}; \nu=1-\nu_{\rm max} \}$
used in (2$\cdot$14) to expand 
$\Psi^{({\rm closed})}_{JM}$ are introduced as follows
\cite{Hiyama2003}:
$\Phi_{JM, \,\nu}$ are written as a sum of the component
functions in the Jacobi coordinate sets $c=1-3$ (Fig.~1),
\begin{equation}
\Phi_{JM, \,\nu}=\Phi_{JM, \,\nu}^{(1)}({\boldsymbol r}_1,
 {\boldsymbol R}_1)
+\Phi_{JM, \,\nu}^{(2)}({\boldsymbol r}_2, {\boldsymbol R}_2)
+\Phi_{JM, \,\nu}^{(3)}({\boldsymbol r}_3, {\boldsymbol R}_3) \; .
\end{equation}
Each component  is expanded in terms of the Gaussian basis 
functions of  ${\boldsymbol r}_c$ and ${\boldsymbol R}_c$:
\begin{equation}
\!\!\!\!\Phi_{JM, \,\nu}^{(c)}({\boldsymbol r}_c, {\boldsymbol R}_c)  
=\!\!\!\sum_{n_cl_c,N_cL_c}A^{(c)}_{J \,\nu,\, n_cl_c,\,N_cL_c} \:
\left[\phi^{\rm G}_{n_cl_c}({\boldsymbol r}_c)\:
\psi^{\rm G}_{N_cL_c}({\boldsymbol R}_c)
\right]_{JM},
(c=1-3)
\label{eq:bases}
\end{equation}
where the Gaussian ranges are postulated to
lie in a geometric progression: 
\begin{eqnarray}
&&\phi^{\rm G}_{nlm}({\boldsymbol r}) =
r^l\:e^{-(r/r_n)^2} \:Y_{lm}({\widehat {\boldsymbol r}}) , \quad
\quad  r_n=r_1\, a^{n-1} , \;  (n=1-n_{\rm max}) 
\label{eq:3gaussa}\\
&&\psi^{\rm G}_{NLM}({\boldsymbol R}) = 
R^L\:e^{-(R/R_N)^2} \:Y_{LM}({\widehat {\boldsymbol R}}) , 
\quad R_N=R_1\, A^{N-1} . \; (N=1-N_{\rm max})
\label{eq:3gauss}
\end{eqnarray}
The basis functions chosen in this way are suitable for describing
both short-range correlations (mainly due to 
nuclear interactions) 
and long-range asymptotic behavior
simultaneously, and therefore 
they are efficient for describing any three-body configuration
(the closed-channel contribution) in the interaction region
of the intermediate stage of reactions. 
The coefficients 
$A^{(c)}_{J \,\nu,\, n_cl_c,\,N_cL_c}$
in (\ref{eq:bases})
and the eigenenergies $E_{J, \nu}$ of $\Phi_{JM, \,\nu}$
are determined by diagonalizing the three-body Hamiltonian $H$
as (\ref{eq:diag}).

This method in which the total Hamiltonian is diagonalized
using the precise Gaussian basis functions has successfully 
been applied to the study of 
various types of three- and four-body systems 
in physics,\footnote{                      
The most precise three-body 
calculation among the applications is the 
determination \cite{Kino2003,Hiyama2003} of
the antiproton mass\cite{PDG2006}
by analyzing CERN's laser spectroscopic data
on the antiprotonic helium atom, 
${\rm He}^{++}+e^- + \bar{p}$ \cite{Hori}.
The energies of the three-body atom with $J\sim 35$ were
calculated  with an accuracy of 10 significant figures.}
which is reviewed in Ref.~\citen{Hiyama2003}. 

\vskip 0.2cm
i) {\it Basis set for} $\alpha+d+X^-$

\noindent   
   In the calculation for $J=0$, we took $l_c=L_c=0, 1, 2$ and
$n_{\rm max}=N_{\rm max}=15$ for $c=1-3$.   
The total number of the three-body Gaussian basis 
$\left[\phi^{\rm G}_{n_cl_c}({\boldsymbol r}_c)\:
\psi^{\rm G}_{N_cL_c}({\boldsymbol R}_c)
\right]_{JM}$
used to construct the eigenfunctions $\{\Phi_{JM, \,\nu}\}$
amounts to $\nu_{\rm max}=2025$, 
which was found to be sufficiently large for the
present calculation.  For the Gaussian ranges,
we took \{$r_1, r_{n_{\rm max}}, R_1,  R_{N_{\rm max}}\}$
$= \{0.5, 15.0, 1.0, 40.0 \,{\rm fm} \}$, 
which are sufficiently precise
for the present purpose. The 
expansion
(2$\cdot$14) converges quickly 
with  increasing $\nu$, and $\nu \lsim 100 
\; (E_{J \nu} \lsim 1$ MeV above 
the $(\alpha X^-)-d$ threshold) is sufficient for the conversion.
Basis sets for $J \geq 1$ are not shown   
since the contribution 
to the cross section from $J \geq 1$ is  
3 orders of magnitude smaller than that from $J=0$.

\vskip 0.2cm
ii) {\it Basis set for} $\alpha + t\, (^3{\rm He})+X^-$

\noindent  
Since  $l_2=1$ in the exit channel of (2$\cdot$4) and (2$\cdot$5),
the basis $\left[\phi^{\rm G}_{n_cl_c}({\boldsymbol r}_c)\:
\psi^{\rm G}_{N_cL_c}({\boldsymbol R}_c)
\right]_{JM}$
for $J=1$ with \{$(l_1=0, L_1=1)$,
 $(l_2=1, L_2=0)$\} is the most important (see Figs.~3 and 4). 
The next most important set for $J=1$ is 
\{$(l_1=1, L_1=0), (l_2=1, L_2=0)$\}.
We take Gaussian ranges  
$\{r_1, r_{n_{\rm max}}, R_1, R_{N_{\rm max}}\}$ 
$=\{0.4, 15.0, 0.8, 30.0 \, {\rm fm}\}$ with
$n_{\rm max}=N_{\rm max}=12$.
The contribution of  $J=0$ to the $S$-factor is
several times smaller than that of $J=1$; we take
the set $(l_c=L_c=0, 1)$ for $c=1-3$.

\subsection{Result for $^6${\rm Li} production}

In this subsection we study 
the $^6$Li production reaction
(2$\cdot$2) that is the most important CBBN reaction.
A three-body calculation of (2$\cdot$2)
was reported
in Ref.~\citen{Hamaguchi} by Hamaguchi {\it et al.},
and therefore we simply recapitulate it here.
The calculated 
astrophysical $S$-factor $S(E)$ is shown
in Fig.~2 together with the Gamow peak 
at $E_0=33$ keV for $T_9=0.1$ ($kT=8.62$ keV)
around which  $(\alpha X^-)$ is formed.
The contribution from the $s$-wave incoming channel 
with $(l_1=L_1=0) \to (l_2=L_2=0)$  is dominant
and that from the $p$-wave incoming channel 
with $(l_1=0, L_1=1) \to (l_2=0, L_2=1)$
is  3 orders of magnitude smaller.
A significant effect of the closed-channel 
amplitude $\Psi_{JM}^{\rm (closed)}$ in (2$\cdot$12),
which represents the contribution of the 
three-body degree of freedom in the interaction region,
can be seen as follows.
If the term $\Psi_{JM}^{\rm (closed)}$
is omitted in the three-body calculation,
$S(E)$ becomes nearly 3 times smaller than
that in Fig.~2.

The CBBN $S$-factor in Fig.~2 is enhanced  by 
a factor of  $\sim 10^7$ 
compared with that of the SBBN reaction (2$\cdot$1).
This confirms the large enhancement ($\sim 10^8$) 
pointed out by Pospelov,\cite{Pospelov2007} although
his scaling relation for the estimation 
is too naive. Note that the  
$S$-factor in Fig.~2 
is of a  similar magnitude to that in
typical nonresonant photonless SBBN reactions
and that the large enhancement ratio
originates simply from the fact that
the SBBN reaction (2$\cdot$1)  is heavily E1-hindered.

The energy dependence of $S(E)$ in Fig.~2 is approximated
by (2$\cdot$27) with
$S(0)=44.6$~keV~b and $\alpha=-0.18$~b.
The reaction rate is then given, using (2$\cdot$28), 
as\footnote{The expression (2$\cdot$36) 
is slightly different from                  
(4$\cdot$3) in Ref.~\citen{Hamaguchi}. The latter is for the vicinity of
$kT=10$ keV ($T_9=0.116$), while the former is valid for
a wider range of $T_9 \lsim 0.2$.}  
\begin{equation}
N_A \,\langle \, \sigma v \,\rangle = 
2.78  \times 10^8 \, T_9^{-\frac{2}{3}} \, 
{\rm exp}\,(- 5.33 \,T_9^{-\frac{1}{3}}) 
( 1 - 0.62\, T_9^\frac{2}{3} - 0.29\, T_9 )
\quad 
{\rm cm}^3 \,{\rm s}^{-1} \, {\rm mol}^{-1}
\end{equation}
for $ T_9 \lsim  0.2$. On the other hand,  since
the SBBN reaction rate of (2$\cdot$1) is given \cite{Angulo1999} as
\begin{equation}
N_A \,\langle \, \sigma v \,\rangle_{\rm SBBN} = 
14.8 \, T_9^{-\frac{2}{3}} \, 
{\rm exp}\,(- 7.44 \,T_9^{-\frac{1}{3}}) 
( 1 + 6.57\, T_9 + \cdot \cdot \cdot )
\quad 
{\rm cm}^3 \,{\rm s}^{-1} \, {\rm mol}^{-1},
\end{equation}
the CBBN/SBBN ratio of the {\it reaction rates}
is $\sim\!10^9$ at $T_9=0.1$. 
This increase of the ratio  from that of the $S$-factors
($\sim 10^7$) is due to the
reduction  of the CBBN Coulomb barrier experienced by the
incoming particle (deuteron) in (2$\cdot$21).

\begin{figure}[bth]
\begin{center}
\epsfig{file=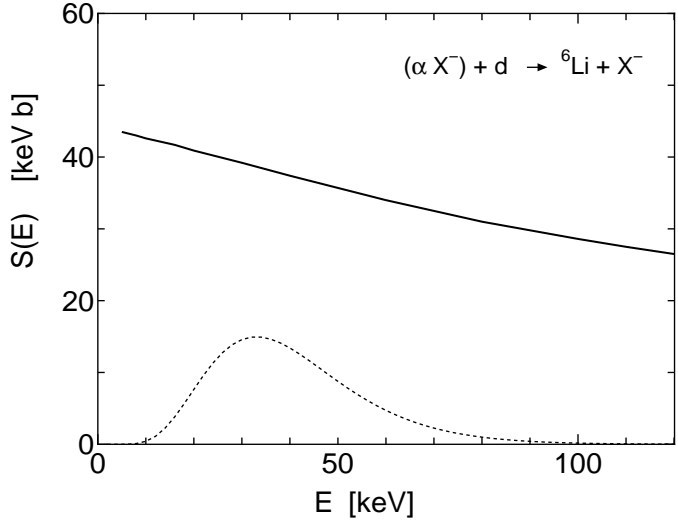,width=9cm,height=7cm}
\end{center}
\caption[]{The 
astrophysical $S$-factor  of the CBBN reaction (2$\cdot$2)
obtained  by the
three-body calculation (solid line). 
The dotted curve (in arbitrary units) 
illustrates the Gamow peak 
for $T_9=0.1$ $(kT=8.62$ keV) with 
the maximum at $E_{\rm 0}=33$ keV.
This figure is taken from Ref.~\citen{Hamaguchi}.
}
\end{figure}


It is desirable to quantitatively examine how sensitive the 
calculated cross section ($S$-factor) is to the choice  of 
the $\alpha$-$d$ potential parameter set
that reproduces the empirical $\alpha$-$d$ binding energy 
and the measured rms charge radius of $^6$Li.
We examined this sensitivity by adopting 
another parameter set \{$r_1=0.9$ fm, $V_1=400$ MeV,
$r_2=2.5$ fm, $V_2=-35.04$ MeV\}, which is markedly
different from the set shown in \S 2.4.
However, we found that the new result for the $S$-factor
differs by only $5-6$\% from that in Fig.~2.
One can expect a similar result for other nonresonant
reactions studied in the present paper, although
such a test is not repeated there.

So far, the mass of the $X^-$ particle, $m_X$, 
has been assumed to be 100 GeV. 
However, we found that 
the calculated $S$-factor 
differs  by only $\sim 0.5$ \%
between  $m_X=100$ GeV and $m_X \to \infty$.  
Therefore, we take $m_X=100$ GeV 
throughout the present paper 
except for \S 3, where a resonant reaction is studied.

Using this large CBBN rate (2$\cdot$36), 
Hamaguchi {\it et al.} \cite{Hamaguchi} 
solved the evolution equation for the $^6$Li abundance
after SBBN has frozen out.
This calculation was performed for various sets of 
values of the assumed 
lifetime $\tau_X$  and the initial number density $n_{X^-}$ of
the $X^-$ particle together with the calculation of
the number density of the bound state $(\alpha X^-)$ 
as a function of the temperature
(cf. Fig.~5 of Ref.~\citen{Hamaguchi}).
For the limiting case of a long lifetime $(\tau_X \gg 1000 $ s),
they obtained $n_{^6{\rm Li}}/n_{\rm B} \simeq
3.7 \times 10^{-5}\, n_{X^-}/n_{\rm B}$, where 
$n_{^6{\rm Li}}$ and $n_{\rm B}$ are the number densities of
$^6$Li and baryons, respectively. 
Therefore, the observational upper bound 
for the $^6 {\rm Li}$ abundance 
$^6 {\rm Li}<6.1\times 10^{-11}$ ($2\sigma$)~\cite{KKM} 
leads to a remarkable bound for the $X^-$ abundance, 
$n_{X^-}/n_{\rm B} < 1.6 \times 10^{-6}$.
The implication of this result for particle physics is
discussed in Ref.~\citen{Hamaguchi}.

\subsection{Result for  $^7${\rm Li} and $^7${\rm Be} production}

The calculated astrophysical $S$-factor of the $^7$Li production
reaction (2$\cdot$4)
is shown in Fig.~3. 
The magnitude of $S(E_0)$ at the Gamow peak energy is
approximately one order smaller than
that of the $^6$Li production reaction
in Fig.~2.
This is because the 
ground state of $^6$Li has $s$-wave ($l_2=0$) angular momentum
of the $d$-$\alpha$ configuration whereas that of 
$^7$Li has $p$-wave $(l_2=1)$ angular momentum
between $t$ and $\alpha$.
The incident wave
with $l_1=L_1=0$ must cause  the angular-momentum
rearrangement to $l_2=L_2=1$ in the exit channel,
and therefore the most effective
\begin{figure}[bth]
\begin{center}
\epsfig{file=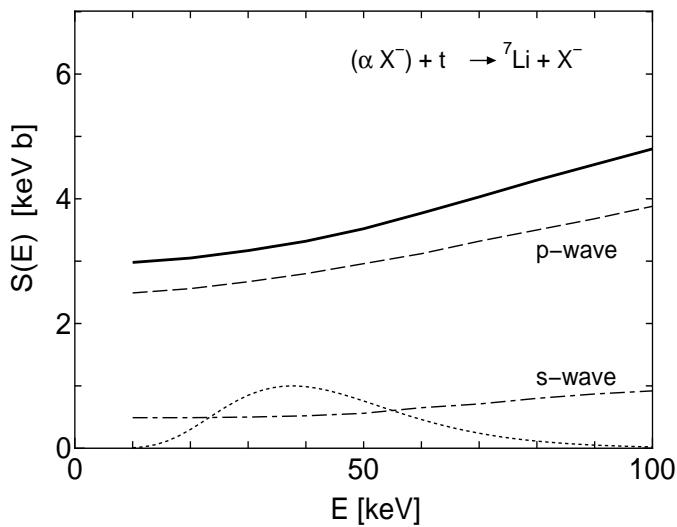,width=9cm,height=7cm}
\end{center}
\caption[]{
The $S$-factor of the CBBN reaction (2$\cdot$4) 
obtained  by the
three-body calculation (the solid line).
The $s$- and $p$-wave contributions are shown individually.
The dotted curve  (in arbitrary units) illustrates
the Gamow peak for $T_9=0.1$ ($kT= 8.6$ keV)
with $E_0=38$ keV.
}
\end{figure}
partial wave is that with $l_1=0, L_1=1$  
in the  incident channel and $l_2=1, L_2=0$ in the exit channel.
This can be seen in Fig.~3; the contribution of the $p$-wave 
$(L_1=1)$ in the incident channel is 
much larger than that of the $s$-wave $(L_1=0)$.

The large effect of the closed-channel 
amplitude $\Psi_{JM}^{\rm (closed)}$ in (2$\cdot$12),
which represents the contribution of the 
three-body degree of freedom in the interaction region
can be seen as follows.
If the term $\Psi_{JM}^{\rm (closed)}$
is omitted in the three-body calculation,
$S(E)$ becomes nearly 50 times smaller than
that in Fig.~3. The effect changes markedly from the case of
$^6$Li production in \S 2.5 (namely, from 3 times
to 50 times). This  is due to the fact that
the angular-momentum transfer
between the entrance and exit channels
is difficult at  low energies much lower than the Coulomb
barrier, and therefore the transfer is 
strongly mediated by the degree of three-body 
distortion in the internal region where the 
reaction takes place.

In the above calculation, the spin of $t$ 
is neglected and therefore the $3/2^-$ ground state and 
$1/2^-$ excited state
are degenerated. However, this assumption was found to work well  
in a more precise calculation with 
the spin of $t$ and the spin-dependent $\alpha$-$t$ interaction
taken into account explicitly.
Namely, the $S$-factors of the transition 
to the ground and excited states
were found to be, respectively, 53\% and 50\% of the $S$-factor by 
the present spin-neglected calculation (the solid curve in Fig.~3)
in the Gamow peak region, and therefore  
their sum (103\%) deviates from the spin-neglected case
by only 3\%, 
although the above respective percentages deviate from  
the numbers 67\% and 33\% (proportional to the spin weights)
in the case where 
the $3/2^-$ ground and $1/2^-$ excited states 
are assumed to have the same degenerated binding energy and 
the same radial wave function.

The magnitude of $S(E)$
for $^7$Li production in CBBN 
is approximately 30 times larger than the
$S$-factor of the E1 radiative capture 
SBBN reaction (2$\cdot$6) in the Gamow peak region  in Fig.~3, 
 $S_\gamma(E_0) \sim 0.1$ keV~b\cite{Angulo1999}.
This CBBN/SBBN enhancement factor of $\sim$ 30 
is very much smaller than that of
$\sim\!10^7$ for $^6$Li production.
This is simply because the E1 transition in SBBN is allowed 
in $^7$Li production but is heavily hindered 
in $^6$Li production. However, as was pointed out in \S 2.1.2.,
Cyburt {\it et al.} \cite{Cyburt2006}
predicted a very large 
CBBN/SBBN enhancement factor ($\sim 10^5$)
assuming the scaling relation  \cite{Pospelov2007}.
This large overestimation is because
the scaling relation, which compares the wavelengths of 
the real and virtual photons,
is not suitable for this type of transfer reaction 
caused by the strong nuclear interaction.
On the other hand, Kusakabe {\it et al.}\cite{Kusakabe2007}
neglected the CBBN processes for the creation of 
$^7$Li and $^7$Be in their BBN network calculation
by taking the same consideration as above 
for the angular-momentum rearrangement 
$(l_1=L_1=0) \to (l_2=L_2=1)$.
Although consideration of the largest contribution from 
$(l_1=0, L_1=1) \to (l_2=1, L_2=0)$ is missing in 
Ref.~\citen{Kusakabe2007}, we support their assumption
as a reasonable one.
However, our reaction rates 
given below will be employed in the BBN network
calculation when
the abundances of $^7$Li and $^7$Be are precisely discussed.

The energy dependence of $S(E)$ in Fig.~4 may be approximated
by expression (2$\cdot$27) with
$S(0)=2.6$~keV~b and $\alpha=0.02$~b, and therefore
the reaction rate is written, for $ T_9 \lsim  0.2$, as 
\begin{equation}
N_A \,\langle \, \sigma v \,\rangle = 
1.4  \times 10^7 \, T_9^{-\frac{2}{3}} \, 
{\rm exp}\,(- 6.08 \,T_9^{-\frac{1}{3}})
(  1 + 1.3\, T_9^\frac{2}{3} + 0.55\, T_9  )
 \quad 
{\rm cm}^3 \,{\rm s}^{-1} \, {\rm mol}^{-1}.
\end{equation}

The calculated  $S$-factor 
for the $^7$Be production reaction (2$\cdot$5),
is shown in Fig.~4. 
Almost the same discussion as that for $^7$Li production 
can be made.
The $S$-factor $S(E)$ in Fig.~4 may be approximated
with $S(0)=13.7$~keV~b and $\alpha=0.01$~b,  then
the reaction rate is written, for $ T_9 \lsim  0.2$, as 
\begin{equation}
N_A \,\langle \, \sigma v \,\rangle = 
9.4  \times 10^7 \, T_9^{-\frac{2}{3}} \, 
{\rm exp}\,(- 9.66 \,T_9^{-\frac{1}{3}}) 
\, (  1 + 0.20\, T_9^\frac{2}{3} + 0.05\, T_9   )
\quad 
{\rm cm}^3 \,{\rm s}^{-1} \, {\rm mol}^{-1}.
\end{equation}

\begin{figure}[bth]
\begin{center}
\epsfig{file=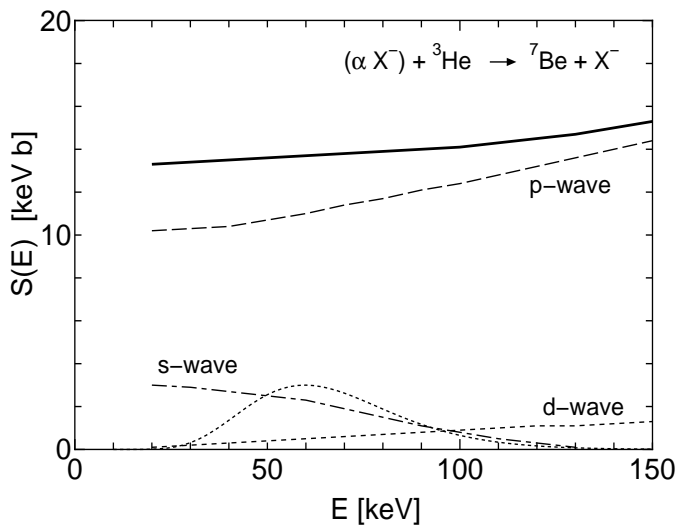,width=9cm,height=7cm}
\end{center}
\caption[]{
The $S$-factor of the CBBN reaction (2$\cdot$5)
obtained  by the three-body calculation  (solid line).
The $s$-, $p$- and $d$-wave  contributions are shown individually.
The dotted curve  (in arbitrary units) illustrates
the Gamow peak for $T_9=0.1$ ($kT= 8.6$ keV)
with  $E_0=60$ keV.
}
\end{figure}


\section{$X^-$-catalyzed radiative capture reactions}

\subsection{Necessity of three-body calculation}

Since the $A=7$ nuclei 
dominantly produced are $^7$Be (eventually decaying to $^7$Li 
by electron capture) in the BBN network calculation
with the CMB-based $\eta_{\rm B}=6.0 \times 10^{-10}$,  
any reaction\footnote{
The reaction $(^7{\rm Be}X^-) + p        
\to\!\!~^8{\rm B}+ X^-$ is 
energetically impossible due to the negative $Q$-value ($-1.323$ MeV);
note that the binding energy of $^8$B is only 0.138 MeV.}
that destroys $^7$Be might be effective
in reducing the overproduction of $^7$Li.
For this purpose,  
Bird {\it et al.} \cite{Bird2007}
considered  a  resonant radiative capture  reaction
(cf. Fig.~5),
\begin{figure}[bth]
\begin{center}
\epsfig{file=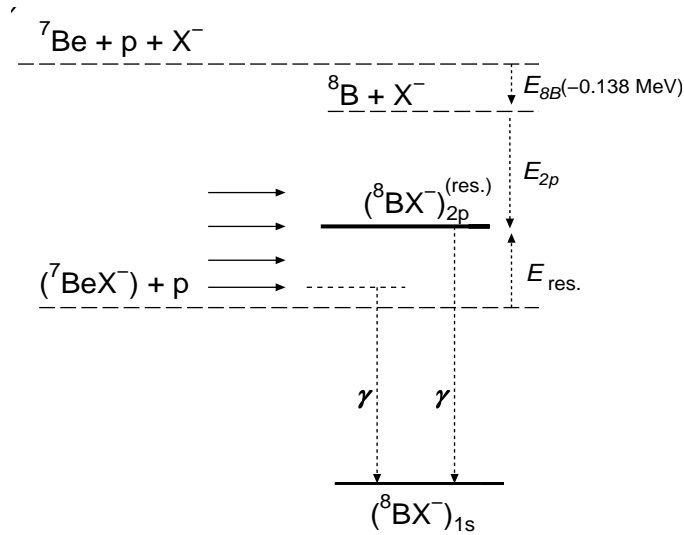,width=9cm,height=7cm}
\end{center}
\caption[]{Schematic illustration of the 
$X^-$-catalyzed 
resonant and nonresonant radiative capture processes
(3$\cdot$1) and (3$\cdot$2).
}
\label{fig:schematic}
\end{figure}
\begin{eqnarray}
&&(^7{\rm Be}X^-) + p \to (^8{\rm B}X^-)_{2\,p}^{\rm (res)}
\to (^8{\rm B}X^-) + \gamma\, (\sim 0.7 \,{\rm MeV})\:,
\end{eqnarray}
 where the  intermediate state, denoted 
as $(^8{\rm B} X^-)_{2\,p}^{\rm (res)}$,
is  a Feshbach resonance generated by the coupling of
the atomic $2\,p$ excited state, say  $(^8{\rm B}X^-)_{2\,p}$,
with the $(^7{\rm Be}X^-) + p$ continuum.
Here, note that after rapid $\beta$-decay, the bound state
 $(^8{\rm B} X^-)$ transforms to  
 $(^8{\rm Be}(2^+, 3\, {\rm MeV}) X^-)$, which immediately
decays\footnote{The authors of             
Refs.~\citen{Bird2007} and \citen{Pospelov2008-9Be} 
erroneously consider that
$(^8{\rm B} X^-)$ transforms to $(^8{\rm Be(gs)} X^-)$
and that the latter state could potentially
lead to a new primordial
source of $^9$Be via  
$(^8{\rm Be} X^-) + n\\
 \to~\!^9{\rm Be}+X^-$.
Note that $^8$B shows no $\beta$-decay to $^8$Be(gs).}
 to the scattering channel $\alpha+ \alpha + X^- + 1.5$ MeV. 

Assuming a simple Gaussian  charge distribution of the
$^8$B nucleus, 
Bird {\it et al.} calculated 
the energy of $(^8{\rm B} X^-)_{2p}$,  $E_{2p}$, 
with respect to the  $^8{\rm B} +  X^-$ threshold.
They obtained $E_{2p}=-1.026$ MeV and estimated
the resonance energy, $E_{\rm res}$,
with respect to the  $(^7{\rm Be} X^-) + p $ threshold
as $E_{\rm res}=
E_{2p}+E_{^8{\rm B}} - E_{(^7{\rm Be} X^-)}
=0.167$ MeV {\it without} the scattering calculation.
Further assuming that  the resonance width 
$\Gamma_{\rm res}$ is much larger than 
the radiative width for decay 
to $(^8{\rm B} X^-)_{1s}$,
they derived the reaction rate of
(3$\cdot$1) and claimed that
the resonant reaction (3$\cdot$1) is 
effective in reducing the  $^7$Li-$^7$Be abundance
in their BBN network calculation  \cite{Bird2007}
and in restricting the lifetime and primordial abundance 
of the $X^-$ particle.

As  discussed below, the rate of the resonant
reaction depends strongly on $E_{\rm res}$, but the
model in Ref.~\citen{Bird2007} is too simple to
treat the resonance state.
Therefore, in this section, we perform a precise 
$^7{\rm Be} + p + X^-$ three-body scattering  calculation
of the resonant reaction (3$\cdot$1) as well as the nonresonant   
radiative capture reaction
\begin{eqnarray}
(^7{\rm Be}X^-) + p  \to (^8{\rm B}X^-) 
+ \gamma \,(\gsim 0.6 \,{\rm MeV})\:,
\end{eqnarray}
although the reaction rate of (3$\cdot$2) 
is much smaller than that of
(3$\cdot$1).

Since the rate of the resonant reaction  
is proportional to exp$(-E_{\rm res}/kT)$,
a small change in  $E_{\rm res}$ can generate a large change
in the rate; for example, 
at $T_9=0.3 \,(kT=25 \,{\rm keV})$,
increase (decrease) of 50 keV in $E_{\rm res}$ 
changes the reaction rate by a factor of
$e^2(=7.4)$. 
The 50 keV change is only 5\% of the binding energy 
($-E_{2p} \sim 1.0$ MeV)
of the $(^8{\rm B} X^-)_{2p}$ state, and therefore
it is essential 
to calculate  $E_{\rm res}$ 
with the detailed structure of the $^8$B 
nucleus taken into account.

Since the nucleus $^8$B has been
extensively studied,  mainly from the viewpoint of the
physics of unstable nuclei and that of the solar neutrino
problem, a large amount of
information about $^8$B has been accumulated
\cite{Tanihata1985,Tanihata1988,Descouv1994,Varga1995,
Csoto1998,Esbenson2004,Ogata2006}.
The ground state of $^8$B, being very weakly bound
($E_{^8{\rm B}}=-0.138$ MeV) 
with respect to the $^7{\rm Be}+p$ threshold,
is known to have a $p$-wave proton {\it halo} (a long-range tail)
with the $^7{\rm Be}+p$ structure. 
Therefore, as will be shown below, the charge distribution of $^8$B  
has a long-range {\it quadrupole} component,
which generates the anisotropic part of the
Coulomb potential between $^8$Be and $X^-$.
The expectation values of this part
for the $(^8{\rm B}X^-)_{2\,p}$ wave function
amount to some $-50$ keV $(J=0)$,
$+10$ keV $(J=1)$ and $+30$ keV $(J=2)$,
where $J$ denotes the total
angular momentum, to which the $p$-wave 
$^7{\rm Be}-p$  relative motion in $^8$B and the $p$-wave
relative motion between $^8$B and $X^-$ 
are coupled (see (3$\cdot$3) below).  

Another important factor that we have to consider 
concerning $^8$B is the spins of the $^7$Be $(3/2^-)$ core 
and the valence proton (1/2). 
The two spins couple with the $p$-wave angular 
momentum between $^7$Be and $p$ 
to give the total angular momentum of $2^+$ in the ground state.
The other possible spin-coupling states 
are not bound; the excitation energies 
of the $1^+$ and $3^+$ resonance
states are 0.77 and 2.32 MeV, respectively. 
Therefore, we cannot neglect 
the fact that the ground state of $^8$B has the specific spin 
of $2^+$.  

However, the estimation of $E_{2\,p}$ in Ref.~\citen{Bird2007}
was made assuming an isotropic Gaussian form of the
charge distribution of $^8$B with neither the $^7{\rm Be}+p$ 
structure nor the spins of the particles taken into account.
In this section, we investigate  
reactions (3$\cdot$1) and (3$\cdot$2),
explicitly adopting the $^7{\rm Be}+p+X^-$ three-body 
degree of freedom in which the  very diffuse, anisotropic 
charge distribution of $^8$B ($=$ $^7{\rm Be}+p)$ is
automatically taken into account. 
The use of this three-body model 
makes it possible to calculate not only  
the precise resonance energy $E_{\rm res}$  but also 
the cross section of the first
transition process in (3$\cdot$1), which could not be  
treated by the model of structureless $^8$B 
in Ref.~\citen{Bird2007}.

\subsection{Resonance energy obtained using approximate  
models of $^8${\rm B}}

Before performing the three-body calculation,
we employ four types of approximate models of $^8$B,
Models i) to iv) below,  and
discuss  how the energy of the 
$(^8{\rm B} X^-)_{2p}$ state, $E_{2p}$, depends on the 
assumed structure of $^8$B, although we do not
couple the state to the $^7{\rm Be}+p$ scattering state.
This may be an instructive guide
to the sophisticated three-body scattering calculation  
in \S \S 3.3 and 3.4.

\vskip 0.2cm
Model i) Bird {\it et al.}~\cite{Bird2007} 
assumed a Gaussian  charge distribution of $^8$B 
with an rms charge radius of $2.64$ fm. 
This gives $E_{2p}=-1.026 $ MeV ($m_X \to \infty$ is taken
throughout this subsection), which 
corresponds to a resonance energy of
$E_{\rm res}^{\rm (i)}=167$ keV above the
$(^7{\rm Be}X^-)+p$ threshold  
at $-1.330$ MeV with respect to the 
$^7{\rm Be}+ X^- +p$ three-body breakup threshold (see Fig.~5).
however, this assumption of the Gaussian charge density
is not appropriate for this special nucleus $^8$B.

\vskip 0.2cm
Model ii) The $^8$B nucleus is known to have a very loosely bound
$^7{\rm Be}+p$ structure with a $p$-wave 
proton halo around the $^7$Be core 
\cite{Descouv1994,Varga1995,Csoto1998,Esbenson2004}. 
Neglecting the spins of  $^7$Be ($3/2^-$) and the 
valence proton ($1/2^-$), 
we firstly calculate the $p$-state $^7{\rm Be}$-$p$ wave function  
using the most well known  $^7{\rm Be}$-$p$ potential 
\cite{Esbenson2004}
(parameters are given in \S 3.3).
By assuming the charge density of $^7$Be (proton) 
to be the Gaussian with the observed rms charge radius
of $2.52$ fm \cite{Tanihata1985} 
($0.8750$ fm \cite{PDG2006}),
we calculate the charge density  
of the ground state of $^8$B. Owing to the $p$-state wave function,
the density has both an isotropic monopole part and 
a deformed quadrupole part.
The former is illustrated in Fig.~6
together with the contributions from 
$^7$Be and $p$.
The density significantly deviates from  
the Gaussian shape in the tail region owing to the 
proton halo.
If we take the $^8$B-$X^-$ Coulomb potential 
due to the monopole charge density alone, 
we have $E^{({\rm mono})}(2p)=-1.009$ MeV, which corresponds to 
$E_{\rm res}^{\rm (ii)}= 184$ keV. 
This increase of $E_{\rm res}$ by 17 keV,
compared with 167 keV in  Model i),
{\it reduces} the reaction rate of (3$\cdot$1)
by a factor of 2 at $T_9=0.3$.

\begin{figure}[htb]
\begin{center}
\epsfig{file=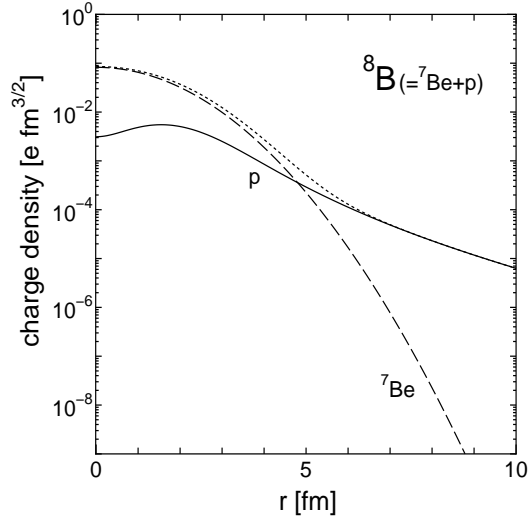,width=7cm,height=7cm}
\end{center}
\caption[]{Monopole part of the charge density, 
$\rho(r)$, of $^8$B (dotted line) calculated using the 
$^7{\rm Be}+p$ model.
The solid (dashed) line shows the contribution from
the proton ($^7$Be). $\rho(r)$ is normalized as
$\int \rho(r) d{\boldsymbol r}=5$, $r$ being the distance
measured from the cm of $^8$B.
}
\end{figure}

\vskip 0.2cm
Model iii) Using the  above  Model ii), we further consider
the contribution from the quadrupole part of the charge density
of $^8$B.
Let $\phi_{1m}({\boldsymbol r}_2)$ 
and $\psi_{1M}({\boldsymbol R}_2)$  denote 
the wave functions of the  
$p$-wave $^8$B ground state  and  the atomic 
$2p$ relative motion between 
$^8{\rm B}$ and $X^-$,  which have 
already been obtained in Model ii).
Then, 
\begin{equation}
\Psi^{{\rm (iii)}}_{JM}=
\left[ \phi_{1}({\boldsymbol r}_2)\otimes\psi_{1}({\boldsymbol R}_2)
\right]_{JM}  \qquad (J=0,1,2)
\end{equation}
is the total wave function
of the $(^8{\rm B} X^-)_{2p}$ 
state\footnote{In this notation       
of $(^8{\rm B} X^-)_{2p}$ in  Model iii),
and similarly in Model iv),
the $p$-wave angular momentum of $^8$B and the total
angular momentum $J$ are not explicitly written for simplicity.}  
with angular momentum $J$ and  $z$-component $M$.
The quadrupole part of the Coulomb 
potential between $^8$B and $X^-$,
which is  sensitive to the angle between ${\boldsymbol r}_2$ and
${\boldsymbol R}_2$,  should contribute to the 
total energy dependently  on
the angular momentum coupling to $J$.
The expectation value of the Hamiltonian 
with respect to $\Psi_{JM}^{({\rm iii})}$,
say $E^{({\rm iii})}_J(2p)$, is written  as
\begin{equation}
E^{({\rm iii})}_J(2p)=E^{({\rm mono})}(2p) + 
\Delta E^{({\rm quad})}_{J}(2p),
\end{equation}
where $\Delta E^{({\rm quad})}_{J}(2p)$ is the contribution from the
quadrupole part of the charge density of $^8$B mentioned above and 
$E^{({\rm mono})}(2p)$ was given in  Model ii).
We obtain $\Delta E^{{\rm (quad})}_J(2p)=-54  \,{\rm keV} \,(J=0)$, 
$+27 \, {\rm  keV}\, (J=1)$ and
$ -5.4 \, {\rm  keV}\, (J=2)$, and therefore,
$E^{\rm (iii)}_J(2p)= -1.063  \,{\rm keV} \,(J=0)$,
$ -1.036 \, {\rm keV}\, (J=1)$ and $-1.014  \,{\rm keV}\, (J=2)$,
which correspond to $E^{\rm (iii)}_{\rm res}= 130 \, {\rm keV}\, (J=0)$,
 $211  \, {\rm keV} \,(J=1)$ and $179 \, {\rm keV}\, (J=2)$.
In the case  of $J=0$, the resonance energy, $E_{\rm res}=130$ MeV,
results in a reaction rate that is 8 times larger than that of 
Model ii).  However, we have not yet
included the effect of the
spins of $^7$Be $(3/2^-)$, $p \,(1/2)$ and $^8$B $(2^+)$.

\vskip 0.2cm
Model iv) The spin-parity 
of the ground state of $^8$B is $2^+$.
In previous studies on $^8$B taking
the $^7{\rm Be}+p$ model, the structure of low-lying states of $^8$B  
is considered as follows:
The valence proton is located in the $p_{3/2}$ orbit around the
$^7$Be ($3/2^-$) core, and the two $3/2^-$ spins
are coupled to the total angular momentum
$I=2^+$ in the ground state and  $I=1^+$ and $3^+$
in the excited states with the observed 
$E_{\rm x}=0.77 $  and 2.32 MeV,
respectively.  Since the spin-dependent interaction is  strong
and the Coulomb interaction between $^8$B and $X^-$ does not 
change the spin structure, we have to seriously consider the 
ground-state spin $2^+$
and its coupling scheme in the calculation of
the resonance energy.
Let $\phi_{\{p_{3/2}, 3/2\} I M_I}({\boldsymbol r}_2)$ 
denote the wave function of
the ground state $(I=2)$ of $^8$B explained above.  
The total wave function
of the $(^8{\rm B} X^-)_{2p}$ state is written as 
\begin{equation}
\Psi^{({\rm iv})}_{JM}= \left[
\phi_{\{p_{3/2}, 3/2\} I }({\boldsymbol r}_2) \otimes 
\psi_1({\boldsymbol R_2}) \right]_{JM},
\end{equation}
where $\psi_1({\boldsymbol R_2})$ was given in  Model ii).
The expectation value of the Hamiltonian 
with respect to $\Psi^{({\rm iv})}_{JM}$,
say $E^{({\rm iv})}_{J,I}(2p)$, is described by the form
\begin{equation}
E^{{\rm (iv)}}_{J,I}(2p)=E^{({\rm mono})}(2p) + \sum_{\Lambda=0}^2\, 
a_{J,I \Lambda} \,\Delta E^{({\rm quad})}_{\Lambda}(2p),
\end{equation}
where $E^{({\rm mono})}(2p)$ and
$\Delta E^{({\rm quad})}_{\Lambda}(2p)$ are given in (3$\cdot$4),
and $a_{J,I \Lambda}$ is written as
\begin{equation}
a_{J,I \Lambda}=\frac{1}{2}(2J+1)(2\Lambda+1) 
\,\sum_{S=1}^2\, [ W(1 I \Lambda J; S 1) ]^2.
\end{equation}
The second term of Eq.~(3$\cdot$6) 
vanishes owing to the summation over $S$ for the ground-state
spin $I=2$ but not for $I \neq 2$.
We then have 
$E^{{\rm (iv)}}_{J,I}(2p)=E^{({\rm mono})}(2p) 
=-1.009$ MeV for $I=2^+$,
and therefore $E^{\rm (iv)}_{\rm res}=184$ MeV $(J=1,2)$
which is, by chance, the same as in  Model ii).

\subsection{Three-body calculation of resonance energy and width}

In the four types of 
approximate models for $(^8{\rm B} X^-)_{2p}$ in the  previous
subsection, $^8$B is assumed to be the same as that in the free space.
However, since $^8$B is a very loosely bound state of
$^7$Be and $p$,
it is possible that the spatial structure of $^8$B changes in the 
presence of $X^-$, namely, in the presence of 
the Coulomb potential
between $^7$Be and $X^-$ and that between $p$ and $X^-$
(the spin structure is not changed by these Coulomb potentials).
Therefore, using the $^7{\rm Be}+p+X^-$ three-body model
with the Hamiltonian (2$\cdot$9), we calculate the resonance state
as a Feshbach resonance embedded in the $(^7{\rm Be} X^-)+p$ continuum
and precisely determine the energy and width of the resonance.

\subsubsection{Nuclear and Coulomb potentials}

In the $^7{\rm Be}+p$ model,
we treat $^7$Be as an inert core with spin $3/2^-$.
We assume Gaussian charge distributions of
$^7$Be and the proton as 
$4e(\pi b_{\rm Be}^2)^{-3/2} e^{-(r/b_{\rm Be})^2}$ 
and $e(\pi b_p^2)^{-3/2} e^{-(r/b_p)^2}$, respectively,  
and take $b_{\rm Be}=2.06$ fm 
and $b_p=0.714$ fm 
to reproduce the observed rms charge radii,
$2.52$ fm \cite{Tanihata1985}  
for $^7$Be and
$0.8750$ fm\cite{PDG2006} for the proton.
The Coulomb potential between $^7$Be and $X^-$ is then given by
\begin{equation}
V_{^7{\rm Be}{\mbox -}X}(r)= -\, 4\, e^2 \, 
\frac{{\rm erf}(r/b_{\rm Be})}{r} ,
\end{equation}
and that between  $X^-$ and $p$ is written as 
\begin{equation}
V_{p{\mbox -}X}(r)= - \, e^2 \, \frac{{\rm erf}(r/b_p)}{r}. 
\end{equation}
The energy of $(^7{\rm Be}X^-)_{1s}$  is 
$ \varepsilon_{\rm gs}^{(1)}=-1.386\,(-1.324)$ MeV
and the rms radius is 3.60 (3.49) fm for $m_X=100 $ GeV
($m_X \to \infty$).

The potential $V_{^7{\rm Be}{\mbox -}p}(r)$ 
is a sum of the nuclear potential,
$V_{^7{\rm Be}{\mbox -}p}^{\rm N}(r)$, 
and the Coulomb potential, $V_{^7{\rm Be}{\mbox -}p}^{\rm C}(r)$.
The latter is given by
\begin{equation}
V_{^7{\rm Be}{\mbox -}p}^{\rm C}(r)= 4 \, e^2
 \, \frac{{\rm erf}(r/\sqrt{b_{\rm Be}^2+b_p^2}\,)}{r} . 
\end{equation}
For the nuclear potential $V_{^7{\rm Be}{\mbox -}p}^{\rm N}(r)$
between $^7$Be and  $p$,
we follow the work of Ref.~\citen{Esbenson2004}, which is 
a standard study on the
$^7{\rm Be}+p \to~\!\!^8{\rm B}+\gamma$ reaction.
The nuclear potential is parameterized as a Woods-Saxon 
potential plus
a spin-orbit potential with an adjustable
depth $V_0(lj,I)$:
\begin{equation}
V_{^7{\rm Be}{\mbox -}p}^{\rm N}(r)=\left[ 1- F_{\rm so} 
({\boldsymbol l} \cdot {\boldsymbol s}) \frac{r_0}{r} \frac{d}{dr} \right]
\frac{V_0(lj,I)}{1+{\rm exp}[(r-R_0)/a]}.
\end{equation}
Here, we take  $a=0.52$ fm,  $R_0=2.391$ fm and 
 $F_{\rm so}=0.351$ fm \cite{Esbenson2004}.
The orbital and total angular momenta
of the proton are denoted by $(lj)$.
The $I=2^+$ ground state of $^8$B is described in terms of 
a pure $(lj)=p_{3/2}$
orbit of the proton coupled with the  $3/2^-$ spin of the
$^7$Be core.
The well depth $V_0~(l=1,j=3/2,I=2)$  is adjusted to reproduce 
the binding energy of 0.1375 MeV of the ground state and is given 
by $-44.147$ MeV.
The $s$-wave potential between $^7$Be and $p$ is assumed to have
$V_0=-30.0$ MeV, $R_0=2.39$ fm and $a=0.65$ fm 
and  additionally have a Pauli repulsive potential with
$V_0=800 $ MeV, $R_0=1.5$ fm and $a=0.2$ fm.
This $s$-wave potential gives no bound state.

\subsubsection{Resonance  wave function}

Let $\xi_{\frac{3}{2}}(^7{\rm Be})$ and
$\xi_{\frac{1}{2}}(p)$ denote the spin functions of
the $^7$Be ground state and the  proton, respectively.
The $1s$ ground state 
of $(^7{\rm Be} X^-)$  is described 
by $\phi_{0 0}^{(1)}({\boldsymbol r}_1)$.  
We consider the $s$-state relative wave function between
the incident proton and the target $(^7{\rm Be} X^-)$,
say $\chi_{00}^{(1)}({\boldsymbol R}_1)
(=\chi^{(1)}_{0}(R_1) Y_{00}(\widehat{\boldsymbol R}_1))$,
since that of the $p$-state has a different parity  
and that of the $d$-state
has negligible contribution to the
resonant radiative capture process.
There are neither inelastically excited channels in $c=1$ 
nor transfer channels in $c=2$ and 3 for the energies
($<200$ keV) concerned here.
The total wave function used to describe the resonant state is then 
written as
\begin{equation}
\Psi_{JM}=\phi^{(1)}_{ 0 0 }({\boldsymbol r}_1)\,
\chi^{(1)}_{00}({\boldsymbol R}_1) 
\left[ \xi_{\frac{3}{2}}(^7{\rm Be})\otimes 
\xi_{\frac{1}{2}}(p) \right]_{JM}
         +\Psi^{({\rm closed})}_{JM}\;.
\end{equation}

The second term of (3$\cdot$12), $\Psi^{({\rm closed})}_{JM}$,
represents the internal amplitude of the 
resonance wave function, whose approximate expression is
$\Psi^{({\rm iv})}_{JM}$ in (3$\cdot$5).
However, we  describe $\Psi^{({\rm closed})}_{JM}$
taking  the three-body degrees of freedom in the same way as 
in \S 2.4. 
Since we find that, in (3$\cdot$12), the contributions
from the amplitudes with $c=1$ and 3 are negligible
in the resonance energy region, we here 
express $\Psi^{({\rm closed})}_{JM}$ as
\begin{equation}
\Psi^{({\rm closed})}_{JM}  =  \sum_{\nu=1}^{\nu_{\rm max}}
  b_{J\,\nu} \; \Phi_{JM,\:\nu}^{(2)}
\end{equation}
using the three-body basis functions only in $c=2$ 
together with the spin functions,
\begin{eqnarray}
\!\!\!\!\!\!\!\!\!\!\!\!\!\!\!\Phi_{JM, \,\nu}^{(2)}  
= \!\! \sum_{n_2,N_2} \!\! A^{(2)}_{J \,\nu,\, n_2,\,N_2} \:
\left[ [\, [\phi^{\rm G}_{n_2l_2}({\boldsymbol r}_2)
\otimes \xi_{\frac{1}{2}}(p)]_{\frac{3}{2}} \otimes
\xi_{\frac{3}{2}}(^7{\rm Be}) ]_I \otimes
\psi^{\rm G}_{N_2 L_2}({\boldsymbol R}_2) \right]_{JM},
\end{eqnarray}
where $l_2=L_2=1, I=2$ and $J=1,2$ are sufficient 
to describe the resonant reaction.
For the Gaussian ranges in (2$\cdot$34) and (2$\cdot$35), we take
$n_{\rm max}=N_{\rm max}=15$ ($\nu_{\rm max}=225$) 
and $\{r_1, r_{n_{\rm max}},R_1, R_{N_{\rm max}}\}=
\{0.4, 15.0, 0.6, 20 \,{\rm fm}\}$,
which is sufficiently precise for the present purpose. 

By diagonalizing the Hamiltonian,
we obtain eigenstates $\{ \Phi_{JM, \nu}; \nu=1-\nu_{\rm max} \}$,
among which the energy of  the lowest-lying  state
($\Phi_{JM,\, \nu=1}$) 
measured from the $^7{\rm Be}+p$ threshold is 
198, 186, 177 and 174 keV for $m_X=50,100$ and 500 GeV and
$m_X \to \infty$, respectively $(J=1,2)$.
The $m_X$ dependence of the energies comes from
the fact that 
the kinetic energy and the $^7{\rm Be}+p$ threshold energy
depend on $m_X$. 

The total wave function (3$\cdot$12) is solved under the
scattering boundary condition
\begin{eqnarray}
\lim_{R_{1}\to\infty} R_1 \chi^{(1)}_{0}(R_1) = 
u^{(-)}_{0}(k_{1},R_{1})
 -S^{J}_{1 \to 1}u^{(+)}_{0}(k_{1},R_{1})
\end{eqnarray}
on the basis of the same prescription as that in \S 2.2.
The resonance state should appear around the energy
of the pseudostate $\Phi_{JM,\, \nu=1}$ mentioned above.
The calculated partial-wave elastic scattering cross section
is illustrated in Fig. 7 for $J=1^-$ with
$m_X=50, 100$ and 500 GeV and $m_X \to \infty$;
similar behaviour is obtained for
the resonance with $J=2^-$. The energy $E_{\rm res}^J$ and
the proton width $\Gamma_p^J$ of the resonance are summarized in 
Table I.

\begin{figure}[htb]
\begin{center}
\epsfig{file=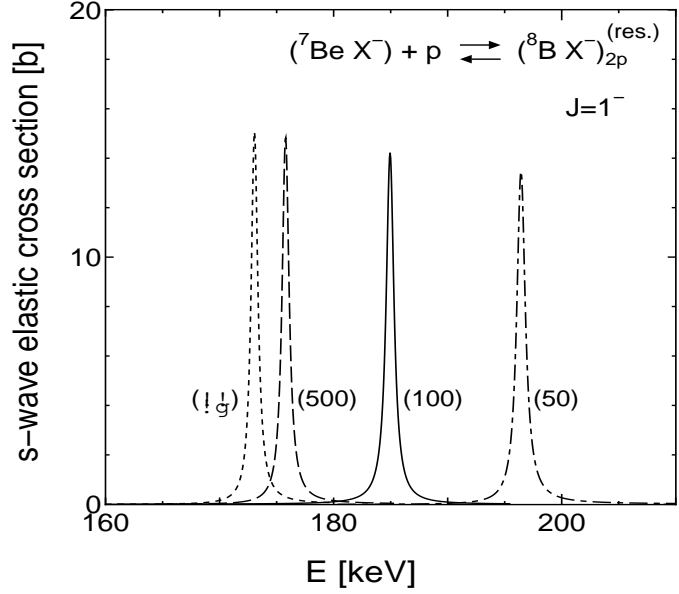,width=9cm,height=8cm}
\end{center}
\caption[]{Partial-wave $(J=1^-$, $s$-wave) cross section
of the elastic $(^7{\rm Be}X)_{\rm 1s}+p$ scattering
near  $(^7{\rm Be} X^-)_{2\,p}^{\rm res}$
for the cases of $m_X=$ 50 GeV (dash-dotted line),
100 GeV (solid line), 500 GeV (dashed line) and 
$m_X \to \infty$ (short dashed line).
Similar behaviour is seen  for $J=2^-$ ($s$-wave). 
}
\label{fig:stau_sigma}
\end{figure}

\subsection{Result for the resonant radiative capture}

The final state, $(^8{\rm B} X^-)$, of the radiative reactions
(3$\cdot$1) and (3$\cdot$2)  is obtained as the ground state
of the $^7{\rm Be}+p+X^-$ system.
The dominant component is obviously 
the product of the $2^+$ ground-state wave function of
$^8$B and the $1s$ wave function of 
$(^8{\rm B}X^-)$. 
We describe it more precisely using the same three-body
basis functions in (3$\cdot$14) with $l_2=1, L_2=0, I=2$ and
$J=2$ with the same ranges of the Gaussian basis.
Other configurations are not necessary within the 
accuracy required for the present purpose.
By diagonalizing the three-body Hamiltonian,
we obtained the ground-state wave function,
$ \Phi_{J=2^+ M}^{({\rm gs})}$ and its energy $E_{\rm gs}$,
which is listed in the seond to last column of Table I.

The width of the electric dipole (E1) transition
from the resonance state $\Psi_{JM}^{\rm (res)}$
with $J^-$ to the $2^+$ ground state 
is given by 
\begin{eqnarray}
\Gamma_\gamma^J &=& \frac{16\pi k_\gamma^3}{9} \frac{1}{2J+1} 
\sum_{M \,M'\,\mu}
|\, \langle \, \Phi_{2 M'}^{\rm (gs)} \,
|\,  Q^{(E1)}_{1 \mu} |\,
\Psi_{JM}^{\rm (res)} \,\rangle \,|^2 \:, \qquad (J=1,2)
\end{eqnarray}
where the E1 operator $ Q^{(E1)}_{1 \mu}$ is defined by  
\begin{eqnarray}
Q^{(E1)}_{1 \mu} &=& \sum_{i=1}^3 q_i R_i^{\rm G} \, Y_{1 \mu}
(\widehat{\boldsymbol R}^{\rm G}_i) \,.
\end{eqnarray}
Here 
$q_i$ and ${\boldsymbol R}_i^{\rm G}$ are the $i$-th particle's
charge and its position vector relative to 
the cm of the total system, respectively, and
$k_\gamma$ is the photon wave number. 
The calculated values of $\Gamma_\gamma^J$ are listed
in Table I; they are, as expected, close to the
width (10.4 eV) given by the simple atomic E1 transition
$(^8{\rm B}X^-)_{2p} \to (^8{\rm B}X^-)_{1s}$.

\begin{table}[th]
\caption{The $^7{\rm Be} + X^- + p$ three-body calculation 
of the energy $(E_{\rm res})$, the proton decay
width $(\Gamma_p$) and the radiative decay width $(\Gamma_\gamma)$
of the resonance states with $J=1^-$ and $J=2^-$
as well as  the 
energy $(E_{\rm gs})$  of the
three-body ground state  with $J=2^+$.
$E_{\rm res}$ and $E_{\rm gs}$  are measured  
from the  $(^7{\rm Be}X^-)_{1s}$ + $p$ threshold, 
whose energy $(E_{\rm th})$ is
given in the last column with respect to the 
three-body breakup threshold. All results are calculated for
$m_X= 50, 100$ and 500 GeV and $m_X \to \infty$. 
}
\label{table:1}
\begin{center}
\begin{tabular}{ccccccccc} 
\hline \hline
\noalign{\vskip 0.2 true cm} 
  &  & $J=1^-$ &(res)  
  &  & $J=2^-$ &(res) 
&   $J=2^+$ (gs) 
 & threshold  
 \\
 &  \multispan3 {\hrulefill} &  \multispan3 {\hrulefill}
  &    &  \cr
\noalign{\vskip 0.1 true cm} 
 $m_X$   
&  $E_{\rm res}^J$    
& $\Gamma_p^J$   
& $\Gamma_\gamma^J$ $\;\;$
& $E_{\rm res}^J$  
& $\Gamma_p^J$ 
& $\Gamma_\gamma^J$   $\;\;$
& $E_{\rm gs}$ 
& $E_{\rm th}$     \\

  [GeV]  
& $\;\;$  [$\,$keV] &  [$\,$keV] &  [$\,$eV]$\;\;$    
& $\;\;$  [$\,$keV] &  [$\,$keV] &  [$\,$eV] $\;\;$   
&  [$\,$keV]   &   [$\,$keV]   \\
\noalign{\vskip 0.2 true cm} 
\hline 
\noalign{\vskip 0.2 true cm} 
  50  
&    196.4  & 0.90  &    9.1 $\;\;$
&    196.9  & 0.55   &   9.1  $\;\;$
&   $-624.2$   &  $(-1252.0)$  \\
  100  
&   185.0 & 0.82   &    9.6 $\;\;$
&  185.5 & 0.50   &    9.6  $\;\;$
&   $-635.5$   &  $(-1286.1)$  \\
  500  
&   175.8  & 0.74   &    9.9 $\;\;$
&   176.3  & 0.44   &    9.9   $\;\;$
&   $-643.6$   &  $(-1316.4)$  \\
  $\infty$  
&    173.0 & 0.71   &    10.1 $\;\;$
&   173.6 & 0.43   &    10.1   $\;\;$
&   $-645.9$   &  $(-1324.0)$  \\
\noalign{\vskip 0.1 true cm} 
\hline\\
\end{tabular}
\label{table:resonance}
\end{center}
\end{table}
%
Finally, the reaction rate of the resonant radiative capture
process is given using the formula for the reaction
(Eq.~(4-194) in Ref.~\citen{Clayton1983}) as 
\begin{eqnarray}
N_A \,\langle \, \sigma v \,\rangle &=&
N_A \hbar^2 \left( \frac{2 \pi}{M_1 k T} 
\right )^{\frac{3}{2}}  
\, \sum_{J=1}^2 \, \frac{2J+1}{(2I_1+1)(2I_2+1)}
\,\frac{\Gamma_p^J \Gamma_\gamma^J}{\Gamma_p^J + \Gamma_\gamma^J} \,
 {\rm exp}\left( -\frac{E_{\rm res}^J}{kT} \right),\nonumber \\
\end{eqnarray}
where $I_1$ and $I_2$ are the spins of 
$^7$Be (3/2) and $p$ (1/2), respectively.  
Here, we consider that the 
total width of the resonance, $\Gamma_{\rm res}$, is given by 
$\Gamma_p^J + \Gamma_\gamma^J$ since there 
is no other decaying channel. We replace the resonance energies
$E_{\rm res}^J (J=1,2)$ by their average, since they
are almost the same. 
We then obtain 
(in units of ${\rm cm}^3 \,{\rm s}^{-1} \, {\rm mol}^{-1}$)
\begin{eqnarray}
N_A \,\langle \, \sigma v \,\rangle &=&
 1.37\times 10^6 \, T_9^{-\frac{3}{2}}\,{\rm exp}\,(-2.28/ T_9), 
\quad (m_X=\;\;50 \:{\rm GeV}) \quad \\
N_A \,\langle \, \sigma v \,\rangle &=&
 1.44\times 10^6 \, T_9^{-\frac{3}{2}}\, {\rm exp}\,(-2.15/ T_9), 
\quad (m_X=100 \:{\rm GeV}) \quad \\
N_A \,\langle \, \sigma v \,\rangle &=&
 1.48\times 10^6 \, T_9^{-\frac{3}{2}}\, {\rm exp}\,(-2.04/ T_9),
\quad (m_X=500 \:{\rm GeV}) \quad \\
N_A \,\langle \, \sigma v \,\rangle &=&
 1.51\times 10^6 \, T_9^{-\frac{3}{2}}\, {\rm exp}\,(-2.01/ T_9) .
\quad (m_X \to \infty ) \quad
\end{eqnarray}

The $m_X$ dependence of the rates is not negligible since,
at $T_9=0.3$, their ratio is  
$0.37 : 0.60 : 0.88 : 1.0 \,$ 
for $m_X= 50, 100$ and 500 GeV and $m_X \to \infty$.
 The rate given by Bird {\it et al.} 
in the second part of Eq.~(3$\cdot$22) in Ref.~\citen{Bird2007}
for $m_X \to \infty$
is by chance close to the above rate (3$\cdot$22);
at $T_9=0.3$, their rate is 1.2 times that of  ours,
although the models are markedly different from each other.

\subsection{Result for the nonresonant radiative capture}

In this subsection, we investigate 
the nonresonant (direct) radiative capture process (3$\cdot$2)
using the $^7{\rm Be}+p+X^-$ three-body model.
Here, we ignore the intrinsic spins of $^7$Be and $p$
because, as will be shown below, the calculated reaction rate
without the spins
is  4 orders of magnitude smaller than that
of the resonant reaction (3$\cdot$1). 
The further inclusion of the spins
will not change the result meaningfully.

We consider the nonresonant E1 radiative capture 
from the $J=0$ incoming state 
of the $(^7{\rm Be}X^-)+p$ channel
to the $J=1$ ground state in which 
the $^7$Be and $p$ are dominantly in  $p$-wave relative motion
(other partial-wave states are negligible),
and we denote the wave functions respectively as
$\Psi_{00}(E)$ and  $\Phi^{\rm (gs)}_{1 M}$
with the proper normalization. 
The wave functions  
can be obtained  using the 
same methods as those in the previous subsections,
but the second term of (3$\cdot$12) can be neglected for the present 
nonresonant scattering wave.

The cross section of the {\rm E1} capture 
is given by
\begin{equation}
\sigma_{\rm cap}^{({\rm E1})}(E)= \frac{16\pi}{9}
\frac{k_\gamma^3}{\hbar v_1}
\sum_M \,
|\, \langle \, \Phi_{1 M}^{\rm (gs)} \,
|\, Q^{(E1)}_{1M} \, |\,
 \Psi_{00}(E) \,\rangle \,|^2 ,
\end{equation}
where 
$k_\gamma$ is the wave number of the emitted photon
and $v_1$ is the velocity of the relative motion between 
$(^7{\rm Be} X^-)$ and $p$.

The calculated $S(E)$ of the CBBN reaction (3$\cdot$2) is
illustrated in Fig.~8  
together with the observed $S$-factor\cite{Angulo1999}
of the SBBN partner reaction 
\begin{equation}
^7{\rm Be}+p \to~^8{\rm B} + \gamma \,(\geq 0.14 \, {\rm MeV}) .  
\end{equation}

\begin{figure}[bth]
\begin{center}
\epsfig{file=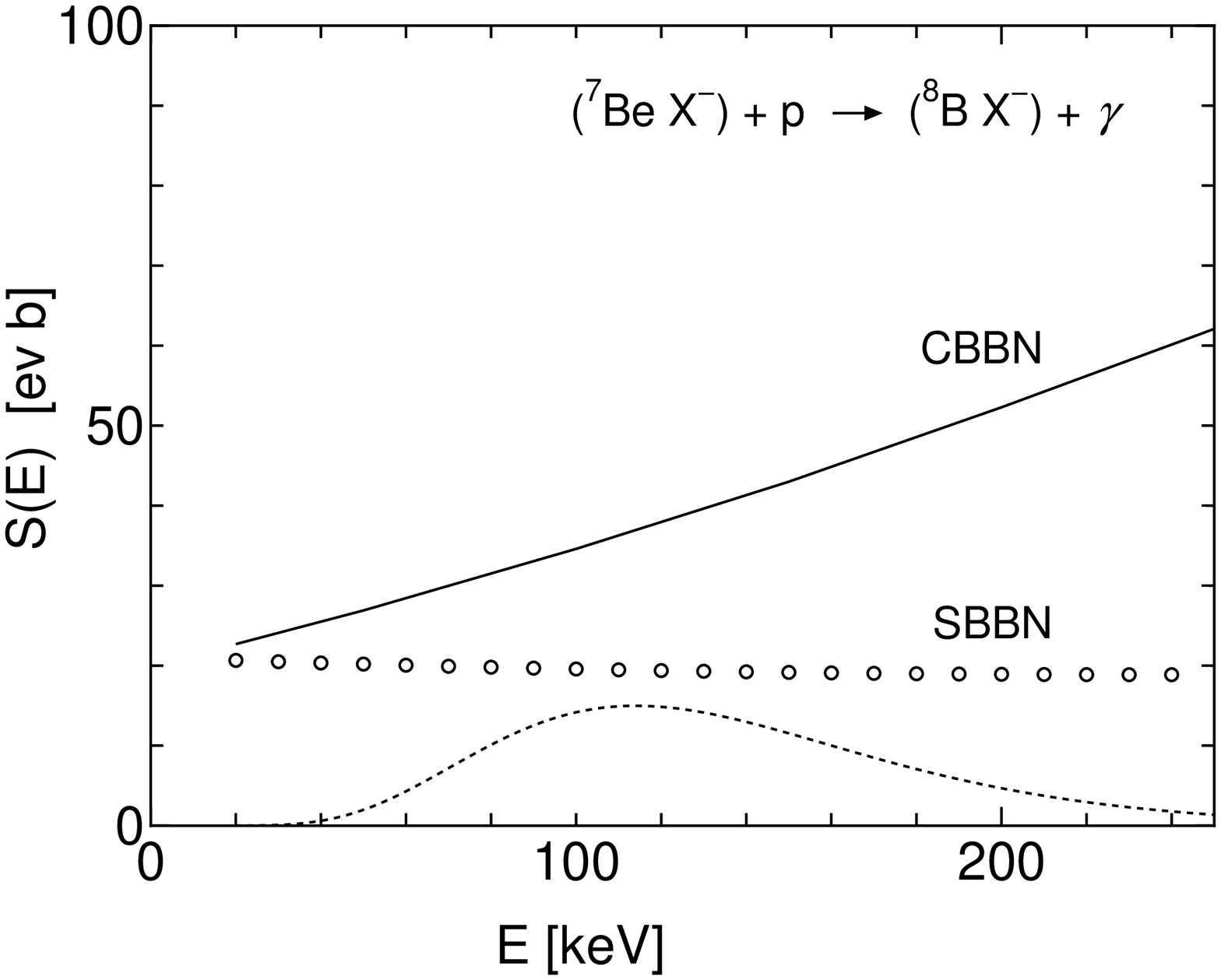,width=8cm,height=6.5cm}
\end{center}
\caption[]{The astrophysical  $S$-factor (solid line) of the
{\it nonresonant} radiative capture CBBN reaction (3$\cdot$2)
obtained by the three-body calculation.
The observed $S$-factor of the SBBN partner reaction (3$\cdot$24)
is shown by  open circles \cite{Angulo1999}.
The dotted curve illustrates
the Gamow peak (in arbitrary units)
for $T_9=0.3$ ($kT=35 $ keV) 
with the  maximum at $E_{\rm 0}=114$ keV.
}
\end{figure}


The enhancement ratio CBBN/SBBN of the $S$-factor 
is only 1 -- 2 in the Gamow peak region 
for the nonresonant radiative 
capture processes (3$\cdot$2) and (3$\cdot$24). 
However,
in the work of Bird {\it et al.} \cite{Bird2007}, 
this ratio is estimated to be as large as $\sim 700$.
The origin of this enormous overestimation 
is the crude assumption employed in their model.
Namely, they assumed that the magnitude of 
the three-body E1 matrix element
in Eq.~(3$\cdot$23) is the same as that of
the two-body E1  matrix element in SBBN.
This is an incorrect assumption because,
in the SBBN matrix element, the dominant contribution comes
from the asymptotic tail region 
of the distance between $^7$Be and the proton (along ${\boldsymbol r}_2$)
in the loosely bound ground state of $^8$B,
but, in the CBBN matrix element, the contribution
is heavily suppressed by the presence of $X^-$ as follows: 
(i) in the $(^7{\rm Be}X^-)_{1s}$ state of the initial channel, 
the distance between $^7$Be and $X^-$ along ${\boldsymbol r}_1$ is 
short (the rms radius is $\sim$ 3.5 fm),
(ii) in the final state $\Psi_{00}(E)$, the distance
between $^8$B and $X^-$ along ${\boldsymbol R}_2$ is 
also short  (the rms radius is $\sim$ 3 fm), and 
thus (iii) these strongly confine 
the possible  $^7{\rm Be}$-$p$ distance 
along ${\boldsymbol r}_2$ that is effective 
in the three-body E1 matrix element. Therefore, the contribution
from the asymptotic region along ${\boldsymbol r}_2$ is heavily suppressed.
The CBBN/SBBN ratio of the $S$-factor
becomes larger in the scaling model of
Ref.~\citen{Bird2007} as the binding energy of $(^8{\rm B}X^-)$
increases, but, at the same time, the wave function of 
$(^8{\rm B}X^-)$ shrinks and therefore the three-body E1 matrix
element becomes smaller. This consideration of this mechanism 
was missing in the simple model of Ref.~\citen{Bird2007} 
for the nonresonant radiative capture reaction.

The fact that the CBBN and SBBN $S$-factors 
have similar magnitudes, within factor nearly 2,
in the Gamow peak region in Fig.~8 
suggests an approximate model
for the cross section $\sigma_{\rm cap}^{({\rm E1})}(E)$ of the
CBBN reaction. The cross section may roughly be derived
from (2$\cdot$21) by employing the observed $S$-factor
of the SBBN-reaction
and the Coulomb barrier penetration factor of the CBBN channel.
Here, the SBBN {\it partner} reaction is defined as
the reaction that is given  
by simply removing $X^-$ from a CBBN reaction\footnote{ 
For instance, the reaction 
$\alpha+d \to~\!^6{\rm Li}+\gamma$ is not
the SBBN partner of $(\alpha X^-)+d 
\to~\!^6{\rm Li}+X^-$ but that of 
$(\alpha X^-)+d 
\to (^6{\rm Li}X^-) + \gamma$.
There is no SBBN¡¡partner of $(\alpha X^-)+d 
\to~\!^6{\rm Li}+X^-$.}.
This approximate model will be further examined in the next section
for complicated CBBN three-body breakup reactions.

In Fig.~8, since the energy dependence of $S(E)$  may be approximated
by (2$\cdot$27) with
$S(0)=20 \times 10^{-3}$~keV~b and $\alpha=0.15 \times 10^{-3} $~b,
the reaction rate is written as 
\begin{equation}
N_A \,\langle \, \sigma v \,\rangle = 
2.3  \times 10^5 \,  T_9^{-\frac{2}{3}} \, 
{\rm exp}\,(- 8.83 \,T_9^{-\frac{1}{3}}) 
\,(  1 + 1.9\, T_9^\frac{2}{3} + 0.54\, T_9   )
\quad 
{\rm cm}^3 \,{\rm s}^{-1} \, {\rm mol}^{-1},
\end{equation}
for $ T_9 \lsim  0.5$.
Since this rate  
is  4 orders of magnitude smaller than
that of the resonant reaction (3$\cdot$1) at $T_9 = 0.3 - 0.5$,
the nonresonant radiative capture reaction (3$\cdot$2) will 
play a very minor role
in the BBN network calculation.


\subsection{Comment on resonant recombination 
between $^7${\rm B\lowercase{e}} and $X^-$}

The radiative capture processes (3$\cdot$1) and (3$\cdot$2) are
preceded by the
recombination of $^7$Be and $X^-$ to form $(^7{\rm Be} X^-)$.
Bird {\it et al.} \cite{Bird2007} calculated 
the reaction rates of the
resonant and nonresonant processes of the recombination.
We here make an important comment on a part of their calculation.
In addition to the normal recombination processes (2$\cdot$7) 
and (2$\cdot$8) in 
Ref.~\citen{Bird2007}, they specially considered
another resonant recombination process,
 (2$\cdot$12) in Ref.~\citen{Bird2007},
\begin{equation}
^7{\rm Be}_{\frac{3}{2}^-} + X^- \to 
(^7{\rm Be}_{\frac{1}{2}^-}  X^-)_{2s} \to   
(^7{\rm Be}_{\frac{1}{2}^-}  X^-)_{2p} + \gamma \to
(^7{\rm Be}_{\frac{3}{2}^-}  X^-)_{1s} +3 \gamma ,   
\end{equation}
where $^7{\rm Be}_{\frac{3}{2}^-}$ and $^7{\rm Be}_{\frac{1}{2}^-}$
are the ground state and  first excited state
at $E_{\rm x}=0.429$ MeV, respectively.  
Their calculated energy of
the intermediate state $(^7{\rm Be}_{\frac{1}{2}^-}  X^-)_{2s}$
with $m_X \to \infty$ is
49 keV {\it below} the 
$^7{\rm Be}_{\frac{3}{2}^-} + X^- $ threshold, which means that
the intermediate state cannot be a resonance
in (3$\cdot$26). 
Bird {\it et al.}, however, artificially pushed the state
upward to a resonant
energy of +10 keV above the threshold
and derived a large recombination rate, Eq.~(2$\cdot$13) of 
Ref.~\citen{Bird2007}.  Their reason for doing this was that
they expected
`a sizable nuclear uncertainty' such as
a correction for the finite mass of $X^-$,
a larger charge radius of the excited state
and a correction for nuclear polarizability.
The contribution of  the `resonant'
$(^7{\rm Be}_{\frac{1}{2}^-}  X^-)_{2s}$ state is then capable of
enhancing the recombination rate by a factor of a few.
Bird {\it et al.} reported two types of results, 
{\it with} and {\it without}
the $(^7{\rm Be}_{\frac{1}{2}^-}  X^-)_{2s}$ contribution,
when they discussed  
the abundance of $^7$Li-$^7$Be and the lifetime and
abundance of $X^-$ particles. 

To completely remove the above `nuclear uncertainty' from the model, 
we perform
an $\alpha+~\!^3{\rm He}+X^-$ three-body calculation of 
the energy of $(^7{\rm Be}_{\frac{1}{2}^-}  X^-)_{2s}$.
The three-body Hamiltonian is the same as that used in 
\S 2, where we investigated the 
reaction $(\alpha X^-)+\!\!~^3{\rm He} \to~\!^7{\rm Be}+X^-$,
but here we take into account the spins of $^3$He and
$^7$Be using a spin dependent $\alpha$-$^3{\rm He}$
potential to reproduce the energies of 
$^7{\rm Be}_{\frac{3}{2}^-}$ and $^7{\rm Be}_{\frac{1}{2}^-}$.
We find that the rms charge radius differs
by only 0.05 fm between 
$^7{\rm Be}_{\frac{3}{2}^-}$ and  $^7{\rm Be}_{\frac{1}{2}^-}$.
According to the three-body calculation the energy of  
$(^7{\rm Be}_{\frac{1}{2}^-}  X^-)_{2s}$
is $-41$ keV for $m_X \to \infty$, $-20$ keV for $m_X=100$ GeV
and $-2$ keV for $m_X=50$ GeV with respect to the 
$^7{\rm Be}_{\frac{3}{2}^-} + X^- $ threshold.
Therefore, we conclude that the 
$(^7{\rm Be}_{\frac{1}{2}^-}  X^-)_{2s}$
state never becomes a resonance in the process of (3$\cdot$26).
In Ref.~\citen{Bird2007}, two types of calculations 
of the element abundance {\it with} and {\it without} the 
$(^7{\rm Be}_{\frac{1}{2}^-}  X^-)_{2s}$
resonance are reported, but the case {\it with} the resonance
is not acceptable.

The application of the 
$\alpha +~\!^3{\rm He}+X^-$  three-body calculation
to the full recombination processes of $^7$Be and $X^-$
will be one of our future subjects of study
since the transition between
the $^7{\rm Be}_{\frac{3}{2}^-}$ and $^7{\rm Be}_{\frac{1}{2}^-}$
states due to $X^-$ is an important factor in the recombination
and  can be unambiguously  treated using the
three-body model.


\section{$X^-$-catalyzed three-body breakup reactions}

The most effective reactions for destroying  $^6$Li and $^7$Li
in SBBN are 
\begin{eqnarray}
&&^6{\rm Li} +\!\!~p \to~\alpha +~\!\!^3{\rm He}
+ 4.02 \,{\rm MeV} \: ,  
\\
&&^7{\rm Li} +\!\!~p \to~\alpha +~\!\!\alpha  
+ 17.35 \,{\rm MeV} \:,   
\end{eqnarray}
which have  large $S$-factors 
($S(0)\sim 3 $ MeV~b and $\sim 0.06$ MeV~b, 
respectively \cite{Angulo1999}). Therefore,
the corresponding CBBN three-body breakup reactions,
\begin{eqnarray}
&&(^6{\rm Li}X^-) +\!\!~p \to~\alpha +~\!\!^3{\rm He} + X^- 
+ 3.22 \,{\rm MeV}\: ,  
\\
&&(^7{\rm Li}X^-) +\!\!~p \to~\alpha +~\!\!\alpha + X^- 
+ 16.47 \,{\rm MeV}\: ,  
\end{eqnarray}
should be taken into account in BBN 
calculations\footnote{Since the                   
kinetic energy of the exit channel is large,
the formation of the bound states $^7$Be, $^8$Be, $(\alpha X^-)$
and  ($^3{\rm He}X^-)$ in (4$\cdot$3) and (4$\cdot$4)
is not important and is not
explicitly considered there.}.
However, the explicit calculation of these CBBN processes is 
tedious and difficult
because
(i) the exit channel is that of a three-body breakup,
(ii) at least a four-body model,
 in which 
$^6$Li ($^7$Li) is composed of 
$d+~\!\!\alpha$ ($t+~\!\!\alpha)$, is needed,
(iii)  $ d ( t ) $ in $^6$Li ($^7$Li) 
must be picked up 
by the incoming proton to form $^3$He ($\alpha$) 
and
(iv) it is tedious to  reasonably determine all the three 
spin-dependent nuclear interactions
appearing in  the four-body model.
Instead, a naive approximation often taken in 
BBN network calculations, for instance, in 
Refs.~\citen{Cyburt2006} and \citen{Kusakabe2007}, 
is that the cross section of the CBBN reaction (4$\cdot$3),
and similarly that of  (4$\cdot$4),
may be given by a product of
the observed $S$-factor of the SBBN partner reaction
(4$\cdot$1) and the CBBN Coulomb barrier penetration factor,
in which the charge of the target is reduced by one unit
owing to the presence of $X^-$.

In this section, we propose 
a more sophisticated
and phenomenologically reasonable three-body model
to implement the 
information of the SBBN cross section
into the CBBN calculation instead of carrying out 
an explicit calculation of (4$\cdot$3) and (4$\cdot$4). 
Firstly, we calculate the SBBN reactions (4$\cdot$1) and (4$\cdot$2).
We do not explicitly treat the channel coupling between the
entrance and exit channels. Instead, we employ only
the entrance channel, say $(A X^-) + a$,
 and introduce a complex potential
$V_{A{\mbox -}a}(r)$ between the particles $A$ and $a$ 
(here,  $A=$$^{6,7}{\rm Li}$ and $a=p$):
\begin{equation}
V_{A{\mbox -}a}(r) = V^{\rm (real)}_{A{\mbox -}a}(r) 
+ i V^{\rm (imag)}_{A{\mbox -}a}(r)   
\end{equation}
as seen in nuclear optical model potentials.
In this case, the absorption cross section
in the elastic $A+a$ scattering
is equivalent to the reaction cross section 
because there are no open channel other 
than the entrance and exit channels of (4$\cdot$1) and (4$\cdot$2).
We determine the potential $V_{A{\mbox -}a}(r)$ so as to
reproduce  the observed SBBN cross section ($S$-factor).

Secondly, we incorporate this potential $V_{A{\mbox -}a}(r)$ 
into the three-body Hamiltonian (2$\cdot$9) of the $A + a +X^-$ system
and solve the elastic scattering between $(A X^-)$ and $a$, 
namely the elastic  $(^{6,7}{\rm Li} X^-) + p$ scattering.
We consider that the absorption cross section obtained 
in this scattering calculation provides the 
cross section of the  CBBN reaction\footnote{   
More precisely, this absorption cross section,
for instance, in the case of $(^6{\rm Li}X^-) +p$ scattering,
includes  transitions to the 
$(\alpha X^-)+~\!\!^3{\rm He}$, $(^3{\rm He} X^-)+ \alpha$
and $^7{\rm Be}+ X^-$ channels,
which are possible in the presence of the $X^-$ particle, but
are of minor importance owing to the much smaller phase space 
than that in the three-body breakup channel.}. 

The  wave function is written similarly to (2$\cdot$12)
but without the exit channel  as
\begin{equation}
\Psi_{JM}=\phi^{(1)}_{00}({\boldsymbol r}_1)\,
\chi^{(1)}_{JM}({\boldsymbol R}_1)   
         +\Psi^{({\rm closed})}_{JM}\;,
\end{equation}
where $\phi^{(1)}_{00}({\boldsymbol r}_1)$ represents 
the $1s$ wave function of  $(^{6,7}{\rm Li}X^-)$ and 
$\chi^{(1)}_{JM}({\boldsymbol R}_1)$ for the
$(^{6,7}{\rm Li}X^-)+p$ scattering wave.  
The scattering boundary condition imposed on  
$\chi^{(1)}_{JM}({\boldsymbol R}_1) ( \equiv \chi^{(1)}_J(R_1)
 Y_{J M}({\widehat {\boldsymbol R}}_1) )$
is given by 
\begin{eqnarray}
\lim_{R_{1}\to\infty} R_1 \chi^{(1)}_{J}(R_1) = 
u^{(-)}_{J}(k_{1},R_{1})
 -S^{J}_{1 \to 1}u^{(+)}_{J}(k_{1},R_{1}) .
\end{eqnarray}

Similarly to in the previous sections, the second term of 
(4$\cdot$6), $\Psi^{({\rm closed})}_{JM}$, represents all the
asymptotically vanishing three-body amplitudes that are
not included in the first scattering term, and is 
expanded in terms of the 
eigenfunctions of the
Hamiltonian (without the imaginary part of (4$\cdot$5))
as
\begin{equation}
\Psi^{({\rm closed})}_{JM}  =  \sum_{\nu=1}^{\nu_{\rm max}}
  b_{J\,\nu} \; \Phi_{JM,\:\nu}     .
\label{eq:Phi-def}
\end{equation}

Using  $S^J_{1 \to 1}$ in (4$\cdot$7),
 we derive the 
reaction cross section as
\begin{equation}
 \sigma_{\rm reac} 
=\frac{\pi}{k_1^2}\sum_{J=0}^{\infty} (2J+1)
        (1- \bigl|S^J_{1 \to 1} \bigr|^2)  .
\end{equation}
This reaction (absorption) cross section
can be expressed alternatively as 
\begin{equation}
\sigma_{\rm reac}=
\frac{-2}{\hbar v_1} \langle \Psi_{JM} \,|  V^{\rm (imag)}_{A-a}(r_2)
\,| \Psi_{JM} \,\rangle .
\end{equation}
These two types of
$\sigma_{\rm reac}$ utilize information from different parts
of the three-body wave function, 
namely, the information
from the asymptotic part along ${\boldsymbol R}_1$ in the former
expression and that from the internal part along  ${\boldsymbol r}_2$ in 
the latter.
Therefore, it is difficult to demonstrate
the agreement between  the two types of
$\sigma_{\rm reac}$.  We obtained
a precise agreement between their numbers 
to four significant figures, which 
demonstrates the high accuracy of  our
three-body calculation.

\subsection{Nuclear complex potential and Coulomb potential}
 
We construct the complex potential\footnote{    
We follow the same prescription
as that employed in Ref.~\citen{Kamimura1989} 
to determine the nuclear fusion rate
in a muonic molecule $(d t \mu)$ by using a
complex potential between $d$ and $t$.}
(4$\cdot$5) between
$^{6,7}$Li and $p$ so as to reproduce the observed 
low-energy $S$-factors of (4$\cdot$1) and (4$\cdot$2) 
(the adopted values in Ref.~\citen{Angulo1999}). 
The shape of the potential is assumed as 
\begin{eqnarray}
 V^{\rm (real)}(r)&=&V_0 \,
\left[1+{\rm exp}\{(r-R_0)/a_0 \}\right]^{-1}
+V^{\rm C}_{{\rm Li}{\mbox -}p}(r) , \\
 V^{\rm (imag)}(r)&=&W_0 \,
\left[1+{\rm exp}\{(r-R_I)/a_I \}\right]^{-1}.
\end{eqnarray}
The Coulomb potential $V^{\rm C}_{{\rm Li}{\mbox -}p}(r)$ 
is derived by assuming a Gaussian
charge distribution for
$^6$Li ($^7$Li) that reproduces the observed rms 
charge radius 2.54 fm (2.43 fm) \cite{Tanihata1985}.
The energies of $(^6{\rm Li} X^-)$ and $(^7{\rm Li} X^-)$
are $-0.773$  and $-0.854$ MeV, respectively,
and the rms radii
are $4.61$ and $4.16$ fm, respectively, when $m_X=100$ GeV.

\vskip 0.2cm
i) $^6{\rm Li} {\mbox -} p$ {\it potential}

The potential parameters are chosen as follows:
$V_0=-59.5$ MeV, $R_0=2.5$ fm  and $a=0.6$ fm
for the odd state and 
$V_0=400$ MeV, $R_0=1.5$ fm and $a=0.2$ fm
for the even state as well as
$W_0=-5.0$ MeV,  $R_I=2.5$ fm and $a_I=0.6$ fm
for both states.
The repulsive potential for the even state is introduced
as a substitute for the Pauli principle to exclude the $s$-state
(the contribution  from the $d$-state is negligible).

\vskip 0.2cm
ii) $^7{\rm Li}{\mbox -} p$ {\it potential}

The potential parameters are chosen as follows:
$V_0=-77.2$ MeV, $R_0=2.5$ fm  and $a=0.6$ fm
for the odd state, 
$V_0=400$ MeV, $R_0=1.5$ fm and $a=0.2$ fm
for the even state, and
$W_0=-1.1$ MeV,  $R_I=2.5$ fm and $a_I=0.6$ fm
for the odd state and $W_0=0$ for the even state.
The reason for setting $W_0=0$ for the even state  is that,
in  reaction (4$\cdot$4), the odd angular momentum between
the two $\alpha$ is prohibited,  
and therefore the even state of the $^7{\rm Li}{\mbox -}p$ 
relative motion is forbidden ($^7$Li has  odd parity).
This explains why the observed $S$-factor of (4$\cdot$4), 
$S(0) \sim 0.06$ MeV~b, is  much smaller than that of
(4$\cdot$3), $S(0) \sim 3$ MeV~b.

\subsection{Result for three-body breakup reactions}

The calculated $S$-factors  
of the CBBN  reactions (4$\cdot$1) and (4$\cdot$2)
are shown in Figs.~9 and 10, respectively, together with 
those of the SBBN partner reactions \cite{Angulo1999}.
It is interesting to note that 
the CBBN/SBBN ratio of the  $S$-factor
is nearly $0.6 -  1$ in the Gamow peak region in 
both figures.  
Also, the ratio was nearly $1 - 2$
in the previous case shown in Fig.~8. 
This suggests  that the CBBN reaction cross section 
may be  approximated, within an error of factor nearly 2, 
by a simple model
in which the CBBN $S$-factor in (2$\cdot$21) is 
replaced by the observed $S$-factor of the SBBN partner 
reaction but the
Coulomb barrier penetration factor 
is kept the same as that in CBBN.
\begin{figure}[bth]
\begin{center}
\epsfig{file=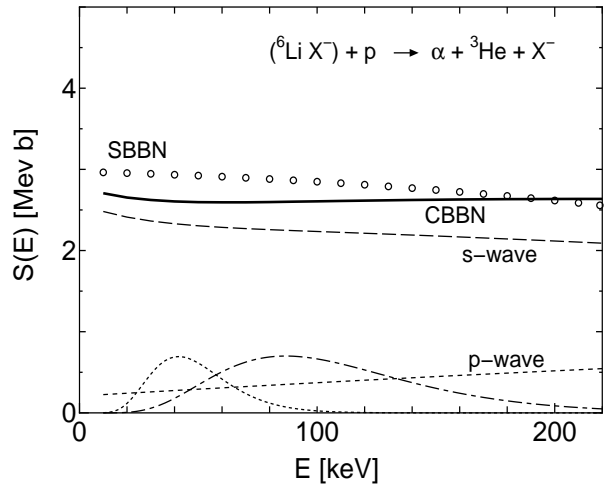,width=8cm,height=6.5cm}
\end{center}
\caption[]{The calculated
$S$-factor for the CBBN reaction (4$\cdot$1)
(solid line).
The $s$- and $p$-wave contributions are shown individually.
The observed $S$-factor of the  SBBN partner reaction (4$\cdot$3)
is given by open circles (the {\it adopted} values in
Ref.~\citen{Angulo1999}).
The dotted curve illustrates
the Gamow peak (in arbitrary units)
for $T_9=0.1$ ($kT=8.6$ keV)  
with the maximum at $E_{\rm 0}=42$ keV,
and the dot-dashed curve is that for $T_9=0.3$ ($kT=26$ keV)  
with $E_{\rm 0}=87$ keV.
}
\end{figure}

This simple model may be used in BBN network calculations
(and has already been employed in some calculations
in the literature)
 when  precise CBBN reaction rate is not   
available. However, the definition of the 
SBBN partner  reaction should be applied strictly
as discussed in \S 3.5.

\begin{figure}[bth]
\begin{center}
\epsfig{file=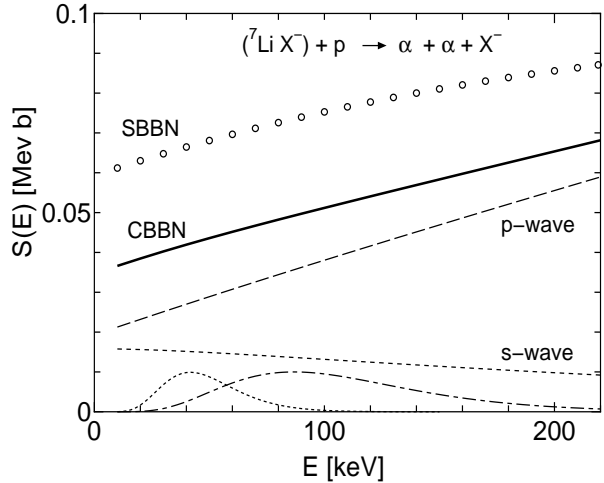,width=8cm,height=6.5cm}
\end{center}
\caption[]{The calculated 
$S$-factor for the CBBN reaction (4$\cdot$2)
(solid line).
The $s$- and $p$-wave contributions are shown individually.
The observed $S$-factor of the SBBN partner reaction (4$\cdot$4)
is given by open circles (the {\it adopted} values 
in Ref.~\citen{Angulo1999}).
The dotted curve is
the Gamow peak (in arbitrary units)
for $T_9=0.1$ ($kT=8.6$ keV)  
with the  maximum $E_{\rm 0}=42$ keV,
and the dot-dashed curve is that for $T_9=0.3$ ($kT=26$ keV)  
with $E_{\rm 0}=87$ keV.
}
\label{fig:stau_fig5}
\end{figure}

The  $S$-factor 
of the CBBN reaction (4$\cdot$1)
in Fig.~9 may be simulated 
using expression (2$\cdot$27) 
with $ S(E) \simeq 2.6 \times 10^3$ keV~b. Using (2$\cdot$28),
the CBBN reaction rate for $T_9 \lsim 0.5$
is then given  as
\begin{equation}
N_A \,\langle \, \sigma v \,\rangle = 
2.6  \times 10^{10} \,T_9^{-\frac{2}{3}} \, 
{\rm exp}\,(- 6.74 \,T_9^{-\frac{1}{3}}) \quad 
{\rm cm}^3 \,{\rm s}^{-1} \, {\rm mol}^{-1} .
\end{equation}

In Fig.~10, which shows  reaction (4$\cdot$2),
the energy dependence of $S(E)$  may be approximated
by (2$\cdot$27) with
$S(0)=36 $~keV~b and $\alpha=0.15 $~b, and therefore
the reaction rate for $ T_9 \lsim  0.5 $  is written as 
\begin{equation}
N_A \,\langle \, \sigma v \,\rangle = 
3.5  \times 10^8 \,
\, T_9^{-\frac{2}{3}} \, 
{\rm exp}\,(- 6.74 \,T_9^{-\frac{1}{3}})
\,(1 + 0.81 \, T_9^\frac{2}{3} + 0.30\, T_9 ) 
 \quad 
{\rm cm}^3 \,{\rm s}^{-1} \, {\rm mol}^{-1}.
\end{equation}

It will be interesting to see whether, 
in the BBN network calculation
using the above reaction rates, 
the CBBN reactions (4$\cdot$1) and (4$\cdot$2)
destroy significant amount of $^{6,7}$Li,  
since their SBBN partners (4$\cdot$3) and (4$\cdot$4)
are the reactions that destroy $^{6,7}$Li
the most strongly in standard BBN.

\section{$X^-$-catalyzed charge-exchange reactions}

When the cosmic temperature cools to $T_9 \sim 0.01$,
the $X^-$ particle 
begins to form the {\it neutral} bound state\footnote{
The binding energy of $(p X^-)$ is $\sim$ 0.025 MeV 
and the rms radius is
$\sim$ 50 fm.}
$(p X^-)$ if $\tau_X \gsim  10^6$ s; this is also
true for $(d X^-)$ and $(t X^-)$, 
even though their fractions are very much smaller
(cf., for example, Fig.~2 in Ref.~\citen{Kusakabe2007}).
Owing to the {\it lack} of a Coulomb barrier,
the bound states are expected to interact strongly
with other elements that have already been 
synthesized.

In this section,  we perform a fully quantum three-body calculation of
the following CBBN reactions that are
induced by the neutral bound states
$(p X^-), (d X^-)$ and $(t X^-)$:

a)  Charge-exchange reactions, 
\begin{eqnarray}
&&(p X^-) + \alpha \to~(\alpha X^-)_{3s,3p,3d} + p \\ 
&&(d X^-) + \alpha \to~(\alpha X^-)_{2s,2p} + d \\
&&(t X^-)\, + \alpha \to~(\alpha X^-)_{2s,2p} +\, t 
\end{eqnarray}

b) Reactions that produce $^{6,7}$Li,
\begin{eqnarray}
&&(d X^-) + \alpha \to~^6{\rm Li} + X^- + 1.4 \:{\rm MeV} \: , \\
&&(t X^-) \, + \,\alpha \to~^7{\rm Li} + X^- + 2.4 \:{\rm MeV} \:, 
\end{eqnarray}

c) Reactions that destroy $^{6,7}$Li and $^7$Be,
\begin{eqnarray}
&&(p X^-)\! +^6\!{\rm Li}\! \to\alpha\! +^3\!{\rm He}\! +\! X^-\! +\!
 4.0 \,{\rm MeV} \:, \\
&&(p X^-)\! +^7\!{\rm Li}\! \to\alpha\! +\alpha +\! X^-\! +\!
 17 \,{\rm MeV}  \:, \\
&&(p X^-)\! +^7\!{\rm Be}\! \to~\!^8{\rm B}\! +\! X^-\! +
 0.11 \,{\rm MeV} \: .
\end{eqnarray}
In the charge-exchange reactions, 
the excited states
of the $(\alpha X^-)$ atom with the principal quantum numbers
$n=3$ and 2 are respectively employed
in (5$\cdot$1) and (5$\cdot$2)$-$(5$\cdot$3).
They have the minimum  transferred energy
among the energetically possible $(\alpha X^-)_{n l}$ 
states.\footnote{The energies of the $1s$ state of    
$(p X^-), (d X^-)$ and $(t X^-)$ are $-25$, $-48$
and $-71$  keV, respectively, whereas those 
of $(\alpha X^-)_{nl}$ are
$-337$ keV$(1s), -96 $ keV$(2p), -90$ keV$(2s), 
-43$ keV$(3d), -43$ keV$(3p)$ and  $-41$ keV$(3s)$.}

Upon performing a DWBA calculation, 
Jedamzik \cite{Jedamzik2007a,Jedamzik2007b}
asserted that
the cross sections of the CBBN reactions (5$\cdot$6) and (5$\cdot$8)
that destroy $^6$Li and $^7$Be, respectively,
are so large that $(p X^-)$ could induce
a second round of BBN, say late-time BBN, capable of
destroying most of the
previously synthesized $^6$Li and $^7$Be.
However, since the strength 
of the nuclear interactions 
(on the order of $\sim 10$ MeV) that cause 
reactions (5$\cdot$6) and (5$\cdot$8) is
much larger than the incident energy $(\lsim 1$ keV),
the Born approximation should not be applied  
to these low-energy nuclear reactions.

Since the neutral bound state
in the entrance channel behaves like a neutron at a large distance
with no Coulomb barrier,
reactions (5$\cdot$4)$-$(5$\cdot$8) 
caused by the nuclear interaction are 
expected to have greatly enhanced cross sections 
even at low energies ($< 1$ keV).
Therefore,
the most interesting issue in this section is whether or not the 
charge-exchange reactions
are so strong that they can intercept   
the reactions (5$\cdot$4)$-$(5$\cdot$8)
that produce and destroy Li and Be.

Before performing a precise quantum  three-body calculation,
we make a  simple  consideration, following the semiclassical 
approach in Refs.~\citen{Pospelov2008-9Be} and \citen{Cohen},
to compare the cross section of the {\it atomic} process 
(5$\cdot$1) with that of the {\it nuclear} 
reactions (5$\cdot$4)$-$(5$\cdot$8). 
In an analogy to the charge-exchange reactions seen in atomic physics,
such as the capture of a muon by a hydrogen atom\cite{Cohen} 
\begin{equation}
(p e^-)_{1s} + \mu^- \to (p \mu^-)_{nl} + e^-, \quad (n \sim 14)
\end{equation}
we understand that,
if the incident $\alpha$ particle in (5$\cdot$1)
enters inside the proton $1s$ orbit of
$(p X^-)$, the sum of the inner charges seen by the proton
becomes positive ($+e$), and therefore the proton 
immediately escapes from the orbit and instead the $\alpha$ particle  
is trapped by $X^-$ as long as
the incident energy is not too high.
In the limit of the semiclassical picture, the cross section of
(5$\cdot$1) is approximately
$\pi b^2$, where  $b$ is the rms radius of 
the $(p X^-)$ atom ($b=50.4$ fm).
On the other hand, 
the cross section of the nuclear reaction 
is roughly $\pi b^2_0$, where $b_0$ is the sum of the radius of the 
incoming nucleus and
the nuclear interaction range ($b_0 \sim $ several fm). 
Therefore, the atomic cross section is roughly two orders of 
magnitude larger than the nuclear cross section.
Actually, this  picture
is not fully applicable because the quantum effect is 
large for low-lying orbits with $n \leq 3$.
However, we find that the following quantum three-body calculations
give a roughly similar ratio of 
the two cross sections, 
although each shows a strong energy dependence
contrary to the above simple consideration.

\vskip 0.2cm
\parindent 10pt
a) Charge-exchange reactions 
  
\parindent 20pt
 The method of calculating the cross sections of  
 the charge-exchange reactions (5$\cdot$1)$-$(5$\cdot$3) 
 is almost the same as that using Eqs.~(2$\cdot$8)$-$(2$\cdot$20).  
 For instance, for the reaction (5$\cdot$1), 
 the entrance channel  $(pX^-) + \alpha$ is 
considered using the 
 Jacobi coordinate system with $c=1$ in Fig.~1, whereas 
 the exit channel $(\alpha X^-)_{nl} + p$ is 
considered using the system with $c=3$.  
 For the closed-channel amplitude $\Psi_{JM}^{\rm (closed)}$ 
 in (2$\cdot$14), we take the same method as that in \S 2.4. 
 The Coulomb potentials are constructed by assuming  Gaussian 
 charge distributions of $p, d, t$ and $\alpha$,  
 the same as in the previous sections. 
 Nuclear interactions play  
 a negligible role in these atomic processes. 
  
 \vskip 0.2cm 
\parindent 10pt
 b) Ractions that produce Li 
  
\parindent 20pt
 The $\alpha$-transfer 
 reactions (5$\cdot$4) and (5$\cdot$5) have the same 
 structure as    
 (2$\cdot$2) and (2$\cdot$4), respectively. 
 Therefore, 
 the method for calculating the cross sections of  
 the former reaction is the same as that in \S 2 except that   
 the entrance channel here is described  using the  
 Jacobi coordinate system with $c=3$ in Fig.~1. 
  
\vskip 0.2cm  
\parindent 10pt
 c) Reactions that destroy Li and Be 
  
\parindent 20pt
 The two reactions (5$\cdot$6) and (5$\cdot$7) have the same 
 structure as  (4$\cdot$3) and (4$\cdot$4), respectively. 
 Thus,  the method for calculating  them   
 is the same.
  The structure of reaction (5$\cdot$8) is similar to 
 that of (3$\cdot$2), although $^8$B and $X^-$ are not in a bound state 
 here but in a photonless scattering state. 
 In (3$\cdot$2), the transition  from the entrance channel to the   
 exit channel via the electromagnetic 
 interaction was treated  perturbatively, but reaction (5$\cdot$8) 
 is solved as a $^7{\rm Be} + p + X^-$ three-body problem. 
  
\parindent 20pt

 \subsection{Results} 
  
 The calculated cross sections $\sigma(E)$  
 of reactions (5$\cdot$1)$-$(5$\cdot$8) are 
 listed in Table~II for energies 
 $E=$0.01, 0.1, 1 and 10 keV $(T_9 \sim 0.0001,  
 0.001, 0.01$ and 0.1, respectively). 
 Since the Coulomb barrier penetration factor exp$(-2\pi \eta(E))$ 
 in (2$\cdot$21) is unity here, the astrophysical $S$-factor 
 is simply given by $S(E)=E \sigma(E)$.

The extreme enhancement   of  the cross sections of   
 reactions (5$\cdot$4)$-$(5$\cdot$8)  due to the 
 neutral bound states  
 can be seen, for example, by the fact that  
 the $^6$Li production cross section of (5$\cdot$4) at $E=10$ keV  
 is $6.4 \times 10^{-1}$ b, whereas 
 that of (2$\cdot$2) with $(\alpha X^-)$ in the entrance channel 
 is $3.9 \times 10^{-6}$ b \cite{Hamaguchi}. 
 As can be seen in the usual {\it neutron}-induced low-energy reactions, 
 the cross sections roughly  follows the  
 $1/v$ law at lower energies ( $\lsim 1 $ keV) in Table II.

 \begin{table}[th] 
 \caption{Calculated cross sections and reaction rates  
 of the late-time BBN reactions induced by the neutral bound states.
 The rates (in units of 
${\rm cm}^3\, 
 {\rm s}^{-1}\, {\rm mol}^{-1}$)
are applicable for  $T_9 \lsim 0.05$. 
 } 
 \begin{center} 
 \begin{tabular}{lllllc}  
 \hline \hline 
 \noalign{\vskip 0.1 true cm}  
& & $\qquad$cross  & $\!\!$section (b) 
 $\;\;$&& reaction  \\ 
 \noalign{\vskip 0.1 true cm}  
$\qquad\qquad$  Reaction &    0.01 {\rm keV}  
 &$ 0.1 ${\rm keV} & 
 $\;1 \,${\rm keV} & $\;$10 \,{\rm keV} & rate
\\  
 \noalign{\vskip 0.1 true cm}  
 \hline  
 \noalign{\vskip 0.05 true cm} 
 a) charge-exchange reaction &  & & \\  
 \noalign{\vskip 0.1 true cm} 
  
 $(p X^-) + \alpha \to~(\alpha X^-)_{3\ell} + p$  & 
 $8.4 \times 10^3  $ & $ 2.2 \times 10^3  $ &  
 $7.8 \times 10^2  $ & $ 7.5 \times 10^1  $ & 
 $1.0 \times 10^{10} $ \\ 
  
 $(d X^-) + \alpha \to~(\alpha X^-)_{2\ell} + d $  & 
 $3.1 \times 10^3 $ & $9.1 \times 10^2 $ &  
 $2.0 \times 10^2 $ & $2.1 \times 10^1 $ & 
 $3.5 \times 10^9 $ \\

 $(t X^-)\, + \alpha \to~(\alpha X^-)_{2\ell} +\, t $  & 
 $8.3 \times 10^3 $ & $1.9 \times 10^3 $ &  
 $3.2 \times 10^2 $ & $2.5 \times 10^1 $ & 
 $7.6 \times 10^9 $ \\ 
  
 \noalign{\vskip 0.15 true cm}

 \hline  
 \noalign{\vskip 0.05 true cm} 
  b) $\alpha$-transfer reaction &  & & \\  
 \noalign{\vskip 0.1 true cm} 
  
 $(d X^-) + \alpha \to~^6{\rm Li} + X^-  $ &  
 $ 9.6 \times 10^0  $ & $ 3.0 \times 10^0 $ &  
 $ 6.9 \times 10^{-1} $ & $ 6.4 \times 10^{-1} $ & 
 $1.1 \times 10^7 $ \\

 $(t X^-) \, + \alpha \to~^7{\rm Li} + X^- $  & 
 $3.5 \times 10^{-1} $ & $1.1 \times 10^{-1} $ &  
 $2.7 \times 10^{-2} $ & $3.0 \times 10^{-2} $ & 
 $4.3 \times 10^5 $ \\ 
  
 \noalign{\vskip 0.15 true cm} 
  
 \hline  
 \noalign{\vskip 0.05 true cm} 
 c) {\rm Li-Be} destruction  & & & & \\  
 \noalign{\vskip 0.10 true cm} 
  
 $(p X^-)\! +^6\!{\rm Li}\! \to\alpha\! +^3\!{\rm He}\! +\! X^-$  & 
 $ 1.8 \times 10^2  $ & $ 5.5 \times 10^1 $ &  
 $ 1.1 \times 10^1  $ & $ 2.8 \times 10^0 $ & 
 $1.6 \times 10^8 $ \\

 $(p X^-)\! +^7\!{\rm Li}\! \to\alpha\! +\alpha +\! X^-$ &  
 $ 3.8 \times 10^{0}  $ & $ 1.7 \times 10^{0}  $ &  
 $ 7.7 \times 10^{-1}  $ & $ 1.6 \times 10^{-1}  $ & 
 $5.5 \times 10^6 $ \\ 
  
 $(p X^-)\! +^7\!{\rm Be}\! \to~\!^8{\rm B}\! +\! X^-$ &  
 $ 4.8 \times 10^{-1}  $ & $ 3.2 \times 10^0  $ &  
 $ 5.3 \times 10^{-1}  $ & $ 5.5 \times 10^{-2} $ & 
 $5.2 \times 10^6 $ \\ 
 \noalign{\vskip 0.15 true cm} 
 \hline\\ 
 \end{tabular} 
 \end{center} 
 \end{table} 
 %

 The most important result in Table II is that 
 the cross sections of the 
 charge-exchange reactions  
 are  much larger than those of the nuclear reactions
 (5$\cdot$4)$-$(5$\cdot$8).  
 Therefore,  for the $\alpha$-transfer reactions 
 (5$\cdot$4) and (5$\cdot$5), 
 the bound states $(d X^-)$ and $(t X^-)$ are   
 mostly changed to $(\alpha X^-)$  
 before producing $^{6,7}$Li.  
 For the Li-Be destruction reactions (5$\cdot$6)$-$(5$\cdot$8), 
 we further note that the probability that  
 $(pX^-)$ comes in contact with $^{6,7}$Li and $^7$Be 
is more than several orders 
 of magnitude smaller than that of it coming in contact with an 
$\alpha$ particle owing to the 
 large difference in the element abundances,  
 and that, even if $(p X)^-$ did collide with  
 an element $A$ (=$^{6,7}$Li or $^7$Be ),  
  another atomic charge-exchange  reaction, 
 \begin{equation}  
 (p X^-) + A \to (A X^-)_{n l} + p \: , \quad 
 (A=~\!^{6,7}{\rm Li}, ^7\!{\rm Be})   
 \end{equation}  
 would immediately occur instead of 
reactions (5$\cdot$6)$-$(5$\cdot$8) 
 since the cross section of (5$\cdot$10) would be as large as  
 that in (5$\cdot$1), although it was not calculated here. 
  
To conclude,  the nuclear reactions (5$\cdot$4)$-$(5$\cdot$8) 
that produce and destroy  
 Li and Be would negligibly change the element abundance 
 in the late-time BBN owing to the strong  
 interception by the charge-exchange reactions 
 (5$\cdot$1)$-$(5$\cdot$3) and (5$\cdot$10). 
  
 We derive the reaction rates of (5$\cdot$1)$-$(5$\cdot$8)  
 for use in the network BBN calculation 
 under the approximation that 
 the cross sections in Table II  
 may be roughly simulated in a simple form (the $1/v$ law) 
 \begin{equation} 
 \sigma(E) \approx C / \sqrt{E}. 
 \end{equation} 
 An evident reduction of the cross section in Table II
from (5$\cdot$11) can be seen at $E \sim 10 $ keV 
 due to the structure of the neutral bound state and 
 the decreased probability of the $\alpha$ particle
being trapped  
 by $X^-$,   but this reduction is thought
to have little effect 
 on the element abundance\footnote{                
 Also, the cross section at $E=0.01$ keV in the bottom line 
 of Table II is much smaller than  that given by 
 (5$\cdot$11), but this is because 
 the $p$-wave contribution is dominant in (5$\cdot$8) 
 and momentum matching becomes  
 difficult at low energies when the proton is 
 transferred to  a definite bound state in the exit channel.}.  
 The use of this approximation in the calculation of 
the reaction rate 
 (2$\cdot$23) makes the result 
 independent of the temperature $T_9$.\footnote{Pospelov 
{\it et al.}\cite{Pospelov2008-9Be} 
derived the rate as $\propto 1/\sqrt{T_9}$, 
but this is due to the fact that they
did not calculate the energy dependence of the cross section
and assumed it to be constant with respect to the energy.
Their value at $T_9=0.01$ is of the same order of magnitude
as ours for $ (p X^-) + \alpha \to (\alpha X^-)_{\rm exc.} + p $.}
 The rate for  $T_9 \lsim 0.05$ is described as
 \begin{equation}  
 N_A \,\langle \, \sigma v \,\rangle   =  
 2.6 \times 10^7 \sqrt{E_{\rm m}/\mu}\: \sigma(E_{\rm m}) 
 \qquad {\rm cm}^3\,{\rm s}^{-1}\, {\rm mol}^{-1} ,  
 \end{equation}  
 where the constant $C$  is determined  
 by matching (5$\cdot$11) to  $\sigma(E)$ in 
 Table II at $E=E_{\rm m}$. Here, $E_{\rm m}$ and  
 $\sigma(E_{\rm m})$ are expressed in units of 
 keV and b, respectively, and $\mu$ is the 
 reduced mass of the entrance channel in units of amu.   
We then simply averaged the three rates determined 
at $E_{\rm m}=0.01, 0.1 $ and 1 (keV).

 The calculated reaction rates are listed in the 
 last column of Table II.
 We consider that it is not necessary to 
 simulate $\sigma(E)$ in a more sophisticated manner
  than (5$\cdot$11),
because it is clear from Table II and the above discussion 
that  the late-time BBN does not affect the  element 
abundances meaningfully\footnote{According to    
 a BBN network calculation  
 by Kusakabe \cite{Kusakabe2008b} 
 including the reaction rates in Table II, 
 the abundances of the bound states $(p X^-), (d X^-)$ and 
 $(t X^-)$ at $T_9 \lsim 0.05$ 
 are greatly reduced by a factor of $\sim 10^5$ 
 from those obtained without including the reaction rates.}. 
  A similar conclusion on the effect of the late-time BBN
was obtained in Ref.~\citen{Pospelov2008-9Be}.


\section{$X^-$-catalyzed production of $^9$B\lowercase{e}} 
  
The absence of stable isotopes with mass number 8 is a
bottleneck for the production of $^9$Be and heavier nuclei in  
standard BBN.
In recent papers \cite{Pospelov2007-9Be,Pospelov2008-9Be},
Pospelov and  coworkers 
pointed out the possibility of the enormous enhancement of
the reaction rate for the production of 
$^9$Be by considering an $X^-$-catalyzed
{\it resonant} neutron-capture process  
\begin{equation}
     (^8{\rm Be} X^-) + n \to (^9{\rm Be}_{\frac{1}{2}^+} X^-)
\to~\!^9{\rm Be}_{\frac{3}{2}^-} + X^-  .
\end{equation}
They claimed that the primordial abundance of $^9$Be
imposes strong restrictions  
on the lifetime and  abundance of $X^-$.
 Here, $^9{\rm Be}_{\frac{3}{2}^-}$ is the 
ground state 
and $^9{\rm Be}_{\frac{1}{2}^+}$ 
 denotes the first excited state with spin ${\frac{1}{2}^+}$ at 
 $E_{\rm x}=1.735 \pm 0.003$ MeV  
 with the neutron decay width  
 $\Gamma_n =0.225 \pm 0.012$ MeV \cite{Sumiyoshi2002} 
 (or $E_{\rm x}=1.684 \pm 0.007$ MeV  
 with $\Gamma_n =0.225 \pm 0.012$ MeV  
 \cite{TUNL2004}). 

In this section, we make a critical comment on 
this work\cite{Pospelov2007-9Be,Pospelov2008-9Be}.
 Their attention to reaction (6$\cdot$1) 
is interesting, but there is 
 a serious problem in their calculation. They assumed  
 that the rms charge radius of both $^8$B 
 and  $^9$Be was 2.50 fm. 
 Since the observed charge radius of  $^9{\rm Be}_{\frac{3}{2}^-}$  
 is $2.519 \pm 0.012$ fm \cite{TUNL2004}, the assumed radius  
 is acceptable for  $^9{\rm Be}_{\frac{3}{2}^-}$. 
 However, there is no reason to take a radius of 2.50 fm  
 for $^8$Be, which is a resonance state at 0.092 MeV  
 above the $\alpha+\alpha$ threshold. 
In principle, the radius of such a resonance state
cannot be experimentally measured, and even theoretically,
it is difficult to define the radius of the oscillating 
resonance wave function.
 Since the width of this resonance is, however, very small (5.57 eV), 
 the wave function is heavily attenuated   
 by the Coulomb barrier, followed by an 
 asymptotically oscillating amplitude 
 that is roughly  three orders of magnitude smaller than that in the 
 nuclear interaction region. 
 Therefore, it is not meaningless to derive the rms radius 
 by using the resonance wave function  
 with the asymptotic part omitted, 
 or by using the wave function obtained through the diagonalization 
 of the Hamiltonian with appropriate $L^2$-integrable basis functions; 
 both methods result in almost the same radius  
 as long as they give the same resonance energy. 
  
 \begin{figure}[bth] 
 \begin{center} 
 \epsfig{file=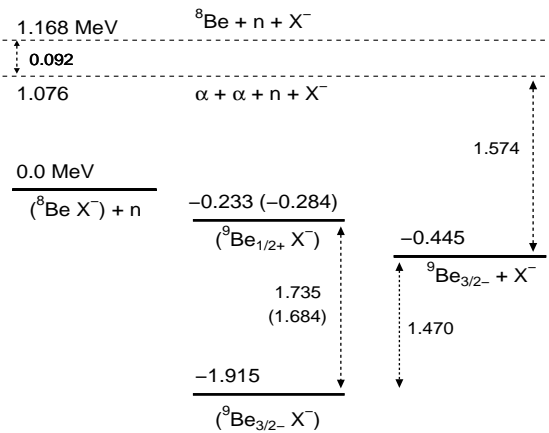,width=8cm,height=7.2cm} 
 \end{center} 
 \caption[]{  
 Isobar diagram for the $^8{\rm Be}+n+X^-$ system. 
 This is to be compared with Fig.~1 of Ref.~\citen{Pospelov2007-9Be}; 
 here the same model as that in Ref.~\citen{Pospelov2007-9Be} 
 is employed but  
 the rms charge radius of $^8$Be is changed to an appropriate value 
 from the too small radius used in Ref.~\citen{Pospelov2007-9Be}(see text). 
 It is demonstrated that   
 the $(^9{\rm Be}_{\frac{1}{2}^+} X^-)$ state 
 becomes {\it below} the   $(^8{\rm Be} X^-)+ n$ threshold  
 contrary to Ref.~\citen{Pospelov2007-9Be}. 
 } 
 \label{fig:jacobi} 
 \end{figure} 

A microscopic $\alpha+\alpha$ model calculation 
 using the resonating group method (RGM) 
 by Arai {\it et al.} \cite{Arai}  
 gives 3.16 fm for the rms charge radius of $^8$Be 
 but the resonance is 0.022 MeV above the $\alpha+\alpha$ threshold.  
 A similar calculation  
 with the orthogonality-condition model (OCM) \cite{Saito} 
 by the present authors and their coworkers
\cite{Hiyamashrink4} 
 gives a radius of 3.39 fm when the $\alpha+\alpha$ resonance   
 is at the observed position of 0.092 MeV.  
 Since the latter model gives 3.20 fm when 
 the resonance energy is tuned to 
the same value (0.022 MeV) as that in the former 
 RGM calculation, we here employ an rms charge radius 
 of $^8$Be of 3.39 fm. 
  
 We now repeat the same analysis as that in 
Refs.~\citen{Pospelov2007-9Be} and \citen{Pospelov2008-9Be}, 
 using the same model but with charge radii of  
 3.39 and 2.52 fm for $^8$Be and $^9$Be, respectively. 
 Here, $m_X \to \infty$ is taken 
similarly to Ref.~\citen{Pospelov2007-9Be}. 
 As  illustrated in Fig.~11, 
 the binding energies of  
 $(^8{\rm Be} X^-)$ and $(^9{\rm Be}_{\frac{3}{2}^-} X^-)$ 
 are respectively 1.168  and 1.470 MeV with respect to 
 the $^8{\rm Be} + X^-$ threshold  
 and  $^9{\rm Be}_{\frac{3}{2}^-}+ X^-$ threshold that is 
 located 1.574 MeV below the $\alpha + \alpha + n + X^-$ 
 threshold. 
 If we assume, similarly to in Ref.~\citen{Pospelov2007-9Be}, 
 that $^9{\rm Be}_{\frac{1}{2}^+}$ has the same charge radius 
 as $^9{\rm Be}_{\frac{3}{2}^-}$, then  
 the $(^9{\rm Be}_{\frac{1}{2}^+} X^-)$ state is 1.735 MeV 
 \cite{Sumiyoshi2002} (or 1.684 MeV \cite{TUNL2004}) above 
 $(^9{\rm Be}_{\frac{3}{2}^-} X^-)$. 
 As a result, we find that  $(^9{\rm Be}_{\frac{1}{2}^+} X^-)$  
 is not a resonance state but a bound state located 
 0.233 MeV (0.284 MeV) {\it below} 
 the   $(^8{\rm Be} X^-)+ n$ threshold  
 with no neutron decay width (although 
 the state has a small width 
 allowing decay to the  
 $^9{\rm Be}_{\frac{3}{2}^-} + X^-$ channel). 
 Those level energies depend 
 on the shape of the charge distributions of $^8$Be and $^9$Be, 
 but the effect on the energy {\it difference} between 
 the  $(^9{\rm Be}_{\frac{1}{2}^+} X^-)$ state and  
 the   $(^8{\rm Be} X^-)+ n$ threshold is very small; 
 for example, the difference changes by only 
 0.005 MeV if the charge distribution is artificially changed from 
 Gaussian to square in both the nuclei. 
 We thus conclude that  
 the resonant neutron-capture process (6$\cdot$1) is not realized 
 as long as the model in Ref.~\citen{Pospelov2007-9Be} is employed. 
  
  \vskip 0.2cm
 Another problem in the analysis in   
 Refs.~\citen{Pospelov2007-9Be} and \citen{Pospelov2008-9Be} is that 
 the charge radius of the excited resonance 
state $^9{\rm Be}_{\frac{1}{2}^+}$  
 is assumed to be the same as that of the ground state of
 $^9{\rm Be}_{\frac{3}{2}^-}$. 
 An $\alpha+\alpha+n$ three-body RGM calculation 
 by Arai {\it et al.} \cite{Arai}  
 gave 2.88 fm for the rms charge radius of  
 $^9{\rm Be}_{\frac{1}{2}^+}$. The calculated energy   
 is $\sim 0.4 $ MeV above the  
 $\alpha+\alpha+n$ threshold  
 while the observed value is   
 $0.161 $ MeV \cite{Sumiyoshi2002} (0.110 MeV \cite{TUNL2004}). 
 The calculated neutron-decay width 
of $\sim 0.2 $ MeV almost reproduces 
 the experimental value. 
 According to Arai {\it et al.} \cite{Arai}, 
an artificial change of the  resonance position, 
 even down to slightly below the  
 $\alpha+\alpha+n$ threshold (by tuning a part of 
the interaction), has little effect on the 
 charge radius of $^9{\rm Be}_{\frac{1}{2}^+}$; 
 this is because the change in the wave function  
is dominantly seen  in the resonating valence neutron 
and little change is seen in the $\alpha$ clusters. 
 If we employ the charge radius of 2.88 fm  
 for $^9{\rm Be}_{\frac{1}{2}^+}$,  
 the energy $-0.233$ MeV ($-0.284$ MeV) of  
 $(^9{\rm Be}_{\frac{1}{2}^+} X^-)$ shown  
 in Fig.~11 must be replaced by $-0.116$ MeV ($-0.167$ MeV). 
 Therefore, the consideration of  
 the differences in the rms charge radii among   
 $^8$Be, $^9{\rm Be}_{\frac{3}{2}^-}$ and  
 $^9{\rm Be}_{\frac{1}{2}^+}$ 
 still causes the $(^9{\rm Be}_{\frac{1}{2}^+} X^-)$ state to lie  
 below the $\alpha+\alpha+n$ threshold.

 In the above discussions on (6$\cdot$1), 
it is still necessary to consider  the change in structure   
 of the {\it loosely coupled} systems  
 $^8$Be and $^9{\rm Be}_{\frac{1}{2}^+}$ 
 induced by the addition of  $X^-$  to them. 
 The dynamics can only be studied 
 by employing an $\alpha+\alpha+n+X^-$ four-body 
 model.\footnote{This type of structure change 
 by the injection of an impurity particle 
 into loosely coupled nuclear states
 has well been studied in light hypernuclei 
by the present authors and their 
coworkers \cite{Hiyamashrink1,Hiyamashrink2,
Hiyamashrink3,Hiyamashrink4}
using three- and four-body models.}
 Use of this model will also make it possible to exactly
 calculate the $(^8{\rm Be} X^-)$ formation processes  
 $ 
 \alpha + X^- \to (\alpha X^-) + \gamma , \;
 (\alpha X^-) + \alpha \to (^8{\rm Be} X^-) + \gamma 
 $ 
 and the resonant (nonresonant) 
 $^9{\rm Be}_{\frac{3}{2}^-}$ formation process  (6$\cdot$1).  
 This study based on the four-body model is beyond  
 the scope of the present paper, 
 but such a study is in progress and will be presented 
 in the near future.

 \section{Summary} 
  
 (1) Using a fully quantum three-body method  developed by the 
 authors\cite{Hiyama2003},
 we have calculated the cross sections of  various types of  
 big-bang nucleosynthesis (BBN) 
 reactions that are catalyzed 
 by a hypothetical long-lived 
 negatively charged, massive leptonic particle (called $X^-$)
 such as the supersymmetric (SUSY) particle {\it stau}. 
 We also provided their reaction rates for use in 
 the BBN network calculation. 
 The rates are summarized in Table III and 
 in the last column of Table II (the late-time BBN reactions
 for $T_9 \lsim 0.05$).

 \vskip 0.2cm 
 (2) In the catalyzed BBN (CBBN) reactions, 
 the strength of the nuclear interaction (on the order of 10 MeV), 
 which causes the transition between the entrance and exit channels, 
 is very much larger than the incident energy ($\lsim 200$ keV), 
 and therefore the coupling between the two channels is 
 sufficiently strong to induce multistep transitions between the channels. 
 Therefore, any calculation method 
 that does not take into account the above property  
 of the interaction is not suitable for application to 
 the reactions. 
 Also, for the charge-exchange reactions (5$\cdot$1)$-$(5$\cdot$3),
 a fully quantum, nonperturbative treatment 
 is highly desirable   
 since the $\alpha$-particle is transferred to low-lying states 
 with the principal quantum number $n= 2 - 3$. 
 We have shown that our three-body calculation method \cite{Hiyama2003} 
 satisfies the above requirements and is useful for the study 
 of all the reactions.

 \begin{table}[th] 
 \caption{Summary of the calculated reaction rates 
of  CBBN reactions obtained by
the three-body calculation.   
 a) - c)  are for  
 $T_9 \lsim 0.2$ 
 and d) - g)  are for $T_9 \lsim 0.5$. 
 } 
 \begin{center} 
 \begin{tabular}{ll}  
 \hline \hline 
 \noalign{\vskip 0.2 true cm}  
 $\qquad\qquad$ Reaction  &  $\qquad\qquad$$\qquad$ 
 Reaction rate (${\rm cm}^3 \,{\rm s}^{-1}\, {\rm mol}^{-1}$) \\  
 \hline  
 \noalign{\vskip 0.05 true cm} 
 $\qquad $ {\it nonresonant} reaction \\ 
 \noalign{\vskip -0.15 true cm} 
 a) $(\alpha X^-) + d \to~^6{\rm Li} + X^- $ & 
 $\quad 
 2.78  \times 10^8 \, T_9^{-\frac{2}{3}} \,  
 {\rm exp}\,(- 5.33 \,T_9^{-\frac{1}{3}})  
 ( 1 - 0.62\, T_9^\frac{2}{3} - 0.29\, T_9 ) $\\ 
 b) $(\alpha X^-) + t \to~^7{\rm Li} + X^- $ & 
 $\quad 
 1.4  \times 10^7 \, T_9^{-\frac{2}{3}} \,  
 {\rm exp}\,(- 6.08 \,T_9^{-\frac{1}{3}}) 
 (  1 + 1.3\, T_9^\frac{2}{3} + 0.55\, T_9  ) $\\ 
 c) $(\alpha X^-) +~\!\!^3{\rm He} \to~^7{\rm Be} + X^- $ & 
 $\quad 
 9.4  \times 10^7 \, T_9^{-\frac{2}{3}} \,  
 {\rm exp}\,(- 9.66 \,T_9^{-\frac{1}{3}})  
 \, (  1 + 0.20\, T_9^\frac{2}{3} + 0.05\, T_9   ) $\\ 
 d) $(^6{\rm Li} X^-) + p \to~\alpha +~\!\!^3{\rm He} + X^- $ & 
 $\quad 
 2.6  \times 10^{10} \,T_9^{-\frac{2}{3}} \,  
 {\rm exp}\,(- 6.74 \,T_9^{-\frac{1}{3}}) $\\ 
 e) $(^7{\rm Li} X^-) + p \to~\alpha +~\!\!\alpha + X^- $ & 
 $\quad  
 3.5  \times 10^8 \, 
 \, T_9^{-\frac{2}{3}} \,  
 {\rm exp}\,(- 6.74 \,T_9^{-\frac{1}{3}}) 
 \,(1 + 0.81 \, T_9^\frac{2}{3} + 0.30\, T_9 )$\\ 
 f) $(^7{\rm Be} X^-) + p \to~(^8{\rm B}X^-)+ \gamma $ & 
 $\quad  
 2.3  \times 10^5 \,  T_9^{-\frac{2}{3}} \,  
 {\rm exp}\,(- 8.83 \,T_9^{-\frac{1}{3}})  
 \,(  1 + 1.9\, T_9^\frac{2}{3} + 0.54\, T_9   ) $\\ 
 \noalign{\vskip 0.20 true cm}  
 \hline  
 \noalign{\vskip 0.05 true cm} 
 $\qquad $ {\it resonant} reaction \\ 
 \noalign{\vskip -0.15 true cm} 
 g) $(^7{\rm Be} X^-) + p \to (^8{\rm B}X^-)_{2p}^{\rm res}$ & 
 $\quad$ $1.37  \times 10^6 \, T_9^{-\frac{3}{2}} \,  
 {\rm exp}\,(- 2.28  T_9^{\,-1}) \qquad   m_X=\;50 {\rm GeV}$ \\ 
  $ \quad \: \to (^8{\rm B}X^-)+\gamma $ & 
 $\quad$ $1.44  \times 10^6 \, T_9^{-\frac{3}{2}} \,  
 {\rm exp}\,(- 2.15  T_9^{\,-1}) \qquad   m_X=100 {\rm GeV}$ \\ 
  & 
 $\quad$ $1.48  \times 10^6 \, T_9^{-\frac{3}{2}} \,  
 {\rm exp}\,(- 2.04 T_9^{\,-1}) \qquad   m_X=500 {\rm GeV}$ \\ 
  & 
 $\quad$ $1.51  \times 10^6 \, T_9^{-\frac{3}{2}} \,  
 {\rm exp}\,(- 2.01  T_9^{\,-1}) \qquad   m_X \to \infty $ \\ 
 \noalign{\vskip 0.15 true cm} 
 \hline\\ 
 \end{tabular} 
 \end{center} 
 \end{table} 

 \vskip 0.2cm 
 (3) In the typical photonless CBBN reactions a)$-$c)  
 in Table III, the calculated  $S$-factors 
 have similar orders of magnitude to those of the 
 typical nonresonant photonless reactions in standard BBN (SBBN). 
 As was first pointed out by Pospelov \cite{Pospelov2007} and then 
confirmed by Hamaguchi {\it et al} \cite{Hamaguchi}, 
 the $S$-factor of CBBN reaction a) that produces $^6$Li is 
 many orders of magnitude larger than that of the 
 radiative capture  $\alpha + d \to\!^6{\rm Li}+\gamma$ 
 in SBBN. This imposes strong restrictions
 on the lifetime and  abundance of  $X^-$.
However, this large enhancement is simply because this SBBN reaction 
 is  heavily E1-hindered.  
 On the other hand, the $S$-factors of  
 CBBN reactions b) and c) are found to be
only $\sim 30$ times larger than 
 those of the strong E1 radiative capture SBBN reactions  
 $\alpha + t (^3{\rm He}) \to\!^7{\rm Li}(^7{\rm Be})+\gamma$. 
 Therefore,  reactions b) and c) 
 do not seem  to change the abundances of $^7$Li-$^7$Be 
 meaningfully.

 \vskip 0.2cm 
 (4) The resonant radiative capture CBBN reaction,  
 g) in Table III (cf. Fig.~5), was proposed by Bird {\it et al.} 
 \cite{Bird2007} as a possible 
solution to the overproduction of $^7$Li-$^7$Be. 
 We have examined their model and result using  
 a $^7{\rm Be}+p+X^-$ three-body model, which makes it possible to 
 calculate all the steps in the reaction. 
 We have shown that
both the very loosely coupled 
 $^7{\rm Be}+p$ structure of $^8$B and 
the strongly spin-dependent $^7{\rm Be}_{\frac{3}{2}^-}-p$  
 interaction  
contribute largely to the energy of the  
 resonance $(^8{\rm B}X^-)_{2p}^{\rm res}$. 
However, the effects work oppositely, almost cancelling 
 each other, and therefore the  
 reaction rate obtained by the simple model \cite{Bird2007}(assuming 
 a Gaussian charge distribution of $^8$B) 
 is by chance close to ours. 
 The reaction rate is sensitive to $m_X$; 
 the rate with $m_X=100$ GeV is 60\% of that with $m_X \to \infty$ 
 at $T_9=0.3$. 
  
 Regarding the recombination process that forms 
 $(^7{\rm Be}_{\frac{3}{2}^-}X^-)$  
 starting from $^7{\rm Be}_{\frac{3}{2}^-}$ and a proton,  
 we pointed out in \S 3.6 that the calculation\cite{Bird2007}
by Bird {\it et al.} of 
 the energy of the `resonance' state 
 $(^7{\rm Be}_{\frac{1}{2}^-}  X^-)_{2s}$  
 is erroneous  owing
 to the lack of consideration of the 
 nuclear structure of  $^7{\rm Be}_{\frac{3}{2}^-}$  
 and  $^7{\rm Be}_{\frac{1}{2}^-}$; 
 the $\alpha +~\!^3{\rm He}+X^-$ three-body calculation 
 clarifed that the  $(^7{\rm Be}_{\frac{1}{2}^-}  X^-)_{2s}$ state 
cannot be a resonance 
 but must be a bound state below 
the $^7{\rm Be}_{\frac{3}{2}^-} + X^- $  threshold.   
 In Ref.~\citen{Bird2007} 
 two types of results were reported on the abundance 
 of $^7$Li-$^7$Be calculated   
 {\it with} and {\it without}  
 the $(^7{\rm Be}_{\frac{1}{2}^-}  X^-)_{2s}$ 
 resonance, but the case {\it with} the resonance 
 is not acceptable.

 \vskip 0.2cm 
 (5) We have performed fully quantum mechanical study on 
 the late-time BBN reactions (5$\cdot$1)$-$(5$\cdot$8),
 which take place  at $T_9 \lsim 0.01$ between the neutral 
 bound states, $(p X^-),$
 $(d X^-)$ and $(t X^-)$, and light 
 nuclei.  The atomic charge-exchange 
 reactions (5$\cdot$1)$-$(5$\cdot$3) are so dominant 
 that they immediately intercept 
 the nuclear reactions (5$\cdot$4)$-$(5$\cdot$8).  Therefore, 
 we conclude that change of element abundances 
 due to the late-time BBN reactions is negligible.  
  
 \vskip 0.2cm 
 (6) We have examined the 'resonant' neutron-capture  
 CBBN process (6$\cdot$1) for producing $^9$Be proposed by  
Pospelov and coworkers\cite{Pospelov2007-9Be,Pospelov2008-9Be}. 
They claimed that the resulting increased output
of $^9$Be strongly constrains 
the abundance and lifetime of $X^-$.   
 However, we found that, 
 as long as the same model as that in Ref.~\citen{Pospelov2007-9Be} 
 is employed but with an appropriate charge radius of  
 $^8$Be (a too small value of the radius was 
 used in Ref.~\citen{Pospelov2007-9Be}), 
 the intermediate state  
 $(^9{\rm Be}_{\frac{1}{2}^+} X^-)$ 
 appears {\it below} the  
 $(^8{\rm Be} X^-) + n $  
 threshold, thus giving no resonant mechanism. 
 The model is, however, too simple to account for the 
 change in the structure of the {\it loosely coupled} (resonant) systems 
 $^8$Be and $^9{\rm Be}_{\frac{1}{2}^+}$ 
 due to the injection of the $X^-$ particle into them. 
 Therefore, an extended  four-body calculation based on 
 an $\alpha+\alpha+n+X^-$
  model  is underway to examine (6$\cdot$1) much more seriously. 
  
 \vskip 0.2cm 
 (7) Finally, we expect that the application of the 
presently obtained reaction rates 
 to the BBN  network calculation will help  solve the 
 $^6$Li-$^7$Li problem and  simultaneously  impose 
 restrictions on the primordial abundance and  lifetime 
of the $X^-$ particle.


\section*{Acknowledgements}
 
 The authors would like to acknowledge helpful discussions
with M. Kusakabe  on the BBN network calculation 
and with K. Arai on the structure of $^8$Be and $^9$Be.
They are also grateful 
 to K. Hamaguchi, T. Hatsuda and T.~T. Yanagida for 
 valuable discussions on  
 stau-catalyzed BBN reactions. 
 This work was supported in part by a Grant-in-Aid for 
 Scientific Research from the Ministry of Education, Culture,
Sports, Science and Technology of Japan. 
 The numerical calculations were perfomed 
 on a HITACHI SR11000 at KEK and a FUJITSU PRIMEQUEST 580 at
Kyushu University.

\end{document}